\documentclass{llncs}

\usepackage{booktabs}
\usepackage{pifont}%
\usepackage{rotating}
\usepackage{amsfonts}
\usepackage{multirow}
\usepackage{color}
\usepackage{graphicx}
\usepackage{hyperref}
\usepackage{amsmath}
\usepackage{algorithm} %
\usepackage{algpseudocode} %
\usepackage{algorithmicx}
\usepackage{flushend}
\usepackage{subcaption}
\usepackage{wrapfig}
\usepackage{breakurl}
\usepackage{enumitem}
\usepackage{xspace}
\usepackage[normalem]{ulem}
\usepackage{listings} %
\usepackage{flushend}
\usepackage{tikz}
\usetikzlibrary{positioning, shapes}
\usepackage{syntax}
\usepackage[version=4]{mhchem} %
\makeatletter
\newcommand{\@chapapp}{\relax}%
\makeatother
\usepackage[title,toc,titletoc,header]{appendix}

\usepackage[disable,textwidth=60pt,textsize=scriptsize]{todonotes} 
\LetLtxMacro{\todom}{\todo}
\renewcommand{\todo}[1]{\todom[inline]{#1}}

\usepackage{relsize}

\usepackage{tabularx}
\usepackage{lipsum}

\newcommand{\DefMacro}[2]{\expandafter\newcommand\csname rmk-#1\endcsname{#2}}
\newcommand{\UseMacro}[1]{\csname rmk-#1\endcsname}

\newcommand{\XComment}[1]{}
\newcommand{\Space}[1]{}

\newcommand{\HighlightCell}[1]{\textbf{#1}}

\definecolor{gray}{RGB}{211,211,211}
\newcommand{\jbasicstyle}{\small\sffamily}

\newcommand{\jnumberstyle}{\scriptsize}

\lstdefinelanguage{pseudo}
{ morekeywords={for, in, break, continue, try, except, not,
  if,else,return,map,fieldElement_array_array40,fieldElement_array40},
  keywordstyle=\bfseries, lineskip=-0.1em, numbers=left,
  numberstyle=\jnumberstyle, numbersep=4pt, basicstyle=\jbasicstyle,
  breaklines=true, breakautoindent=true, tabsize=2,
  columns=fullflexible, morecomment=*[l][\textsl]{//},
  mathescape=true, }
\newenvironment{BNF}
  {\captionsetup{type=lstlisting}}
  {}

\def\figname{Figure}
\def\formulaname{CRN}
\def\algname{Algorithm}

\newfloat{formula}{pb}{form}
\floatname{formula}{\formulaname}

\newcommand{\euler}{e}

\newcommand{\Tool}{\textit{CRN}\nobreak\hspace{-.05em}\raisebox{.4ex}{\bf \textsmaller[3]{+}}\nobreak\hspace{-.10em}\raisebox{.4ex}{\bf \textsmaller[3]{+}}\xspace}
\newcommand{\CRNSimulator}{\mbox{\textit{CRNSimulator}}\xspace}

\newcommand{\xgty}{$X_{gtY}$\xspace}
\newcommand{\xgtyNoSpace}{$X_{gtY}$}
\newcommand{\xlty}{$X_{ltY}$\xspace}
\newcommand{\xltyNoSpace}{$X_{ltY}$}
\newcommand{\ygtx}{$Y_{gtX}$\xspace}
\newcommand{\yltx}{$Y_{ltX}$\xspace}
\newcommand{\xgtyEq}{$X > Y$\xspace}
\newcommand{\xltyEq}{$X < Y$\xspace}
\newcommand{\xeqyEq}{$X = Y$\xspace}

\newcommand{\ldModule}{\textit{ld}\xspace}
\newcommand{\ldModuleCapitalized}{\textit{Ld}\xspace}
\newcommand{\ldModuleNoSpace}{\textit{ld}}
\newcommand{\addModule}{\textit{add}\xspace}
\newcommand{\addModuleCapitalized}{\textit{Add}\xspace}
\newcommand{\addModuleNoSpace}{\textit{add}}
\newcommand{\subModule}{\textit{sub}\xspace}
\newcommand{\subModuleCapitalized}{\textit{Sub}\xspace}
\newcommand{\subModuleNoSpace}{\textit{sub}}
\newcommand{\mulModule}{\textit{mul}\xspace}
\newcommand{\mulModuleCapitalized}{\textit{Mul}\xspace}
\newcommand{\mulModuleNoSpace}{\textit{mul}}
\newcommand{\divModule}{\textit{div}\xspace}
\newcommand{\divModuleNoSpace}{\textit{div}}

\newcommand{\sqrModuleNoSpace}{\textit{sqrt}}
\newcommand{\cmpModule}{\textit{cmp}\xspace}
\newcommand{\cmpModuleCapitalized}{\textit{Cmp}\xspace}
\newcommand{\cmpModuleNoSpace}{\textit{cmp}}

\newcommand{\crnKeywordNoSpace}{\textit{crn}}
\newcommand{\stepKeyword}{\textit{step}\xspace}
\newcommand{\stepKeywordNoSpace}{\textit{step}}
\newcommand{\steps}{steps\xspace}

\newcommand{\ifAbsentKeyword}{\textit{ifAbsent}\xspace}

\newcommand{\gtKeyword}{\textit{ifGT}\xspace}
\newcommand{\gtKeywordNoSpace}{\textit{ifGT}}
\newcommand{\geKeyword}{\textit{ifGE}\xspace}
\newcommand{\geKeywordNoSpace}{\textit{ifGE}}
\newcommand{\eqKeyword}{\textit{ifEQ}\xspace}
\newcommand{\eqKeywordNoSpace}{\textit{ifEQ}}
\newcommand{\leKeyword}{\textit{ifLE}\xspace}
\newcommand{\leKeywordNoSpace}{\textit{ifLE}}
\newcommand{\ltKeyword}{\textit{ifLT}\xspace}
\newcommand{\ltKeywordNoSpace}{\textit{ifLT}}

\newcommand{\crnNonterminal}{Crn}
\newcommand{\rootStatementListNonterminal}{\textit{RootSList}}
\newcommand{\rootStatementNonterminal}{\textit{RootS}}
\newcommand{\concStatementNonterminal}{\textit{ConcS}}
\newcommand{\rxnStatementNonterminal}{\textit{RxnS}}
\newcommand{\moduleStatementNonterminal}{\textit{ModuleS}}
\newcommand{\arithmeticStatementNonterminal}{\textit{ArithmeticS}}
\newcommand{\cmpStatementNonterminal}{\textit{CmpS}}
\newcommand{\stepStatementNonterminal}{\textit{StepS}}
\newcommand{\nestedStatementListNonterminal}{\textit{CommandSList}}
\newcommand{\nestedStatementNonterminal}{\textit{CommandS}}
\newcommand{\conditionalStatementNonterminal}{\textit{ConditionalS}}
\newcommand{\expressionNonterminal}{\textit{Expr}}

\newcommand{\BeginFormula}{\vspace{-20pt}\begin{formula}[H]\vspace{-5pt}}
\newcommand{\EndFormula}{\vspace{-5pt}\end{formula}\vspace{-20pt}}
\newcommand{\Subsubsection}[1]{\vspace{-10pt}\subsubsection{#1}}

\begin{document}

\title{\Tool: Molecular Programming Language}
\author{Marko Vasic \and
  David Soloveichik \and
  Sarfraz Khurshid
}
\institute{The University of Texas at Austin, USA \\
  \email{\{vasic,david.soloveichik,khurshid\}@utexas.edu}
}

\maketitle

\begin{abstract}

Synthetic biology is a rapidly emerging research area, 
with expected wide-ranging impact in biology, nanofabrication, and medicine.
A key technical challenge lies in embedding computation in molecular contexts where electronic micro-controllers cannot be inserted.
This necessitates effective representation of computation using molecular components.
While previous work established the Turing-completeness of chemical reactions,
defining representations that are faithful, efficient, and practical remains challenging.  
This paper introduces \Tool, a new language for programming deterministic (mass-action) chemical kinetics to perform computation. 
We present its syntax and semantics, and build a compiler translating \Tool{} programs into chemical reactions,
thereby laying the foundation of a comprehensive framework for molecular programming.
Our language addresses the key challenge of embedding familiar imperative constructs into a set of chemical reactions happening simultaneously and manipulating real-valued concentrations.
Although some deviation from ideal output value cannot be avoided, we develop methods to minimize the error, and implement error analysis tools.
We demonstrate the feasibility of using \Tool on a suite of well-known algorithms for discrete and real-valued computation.
\Tool{} can be easily extended to support new commands or chemical reaction implementations,
and thus provides a foundation for developing more robust and practical molecular programs.
\end{abstract}
\section{Introduction}

A highly desired goal of synthetic biology is realizing a programmable chemical controller that can operate in molecular contexts incompatible with traditional electronics.
In the same way that programming electronic computers is more convenient at a higher level of abstraction than that of individual flip-flops and logic circuits, we similarly expect molecular computation to admit specification via programming languages sufficiently abstracted from the hardware.
This paper focuses on developing a compiler for a natural imperative programming language to a deterministic (mass-action) chemical reaction network implementing the desired algorithm.
We do not directly make assumptions on how the resulting reactions would be implemented in chemistry.
This could in principle be achieved by DNA strand displacement cascades~\cite{SoloveichikETAL10DNAUniversalSubstrate}, or other programmable chemical technologies such as the PEN toolbox~\cite{baccouche2014dynamic}.

Deterministic (mass-action) chemical kinetics is Turing universal~\cite{FagesETAL17TuringCompletenessOfContinuousCRNs}, 
thus in principle allowing the implementation of arbitrary programs in chemistry. 
Turing universality was demonstrated by showing that arbitrary computation can be embedded in a class of polynomial ODEs~\cite{Bournez16Polynomial}, 
and then implementing these polynomial ODEs with mass-action chemical kinetics.
While these results establish a sound theoretical foundation and show the power of chemistry for handling computation tasks in general,
translating and performing specific computational tasks can lead to infeasibly large and complex sets of chemical reactions.

In this work we develop a programming paradigm for chemistry, based on the familiar imperative programming languages, with the aim of making molecular programming more intuitive, and efficient.
Most commonly used programming languages such as C, Java and Python, are imperative in that they use statements that change a program's state, with typical branching constructs such as if/else, loops, etc.
Note that although CRNs are sometimes talked about as a programming language~\cite{ChenETAL13ProgrammableChemicalControllersFromDNA}, they are difficult to program directly (it is even unfair to equate them with assembly language). In contrast, \Tool operates at a much higher level.

We introduce the syntax and semantics of \Tool, an imperative programming language 
that compiles to deterministic (mass-action) chemical reaction networks.
\Tool has an extensible toolset including a simulation framework and error analysis functionality.
A user specifies a \emph{sequence} of statements, termed commands, to
execute. 
Assignment, comparison, loops, conditional execution, and arithmetic operations are supported.
The generated reactions are logically grouped into modules performing the corresponding command. 
Each module transforms initial species concentrations to their steady-state values which are the output of the module.
We ensure that such modules are composable by preserving the input concentrations at the steady-state.

A mapping of imperative program logic to chemical reactions manipulating continuous concentrations poses various challenges that we must address. 
All reactions happen concurrently, making it difficult to represent sequential computation where, 
for example, the result of one operation is first computed and then used in another operation. 
Similarly, all branches of the program execution (i.e., if / else) are followed simultaneously to some degree. 
To mimic sequential execution in mass-action chemistry, we ensure that the reactions corresponding to the current command happen quickly, 
while other reactions are slow.  
For this we rely on a chemical oscillator in which the \emph{clock} species oscillate between low and high concentrations, and catalyzing reactions with different clock species. 
To achieve conditional execution, we further need to ensure that the reactions corresponding to the correct execution branch happen quickly, while those corresponding to other branches are inhibited. 
Our \cmpModule module sets \emph{flag} species to reflect the result of comparison, and these species catalyze the correct branch reactions.

A further source of error is that the set of basic modules, such as addition, converge to the correct value only in the limit, 
thus computing approximately in finite time. 
To mitigate this source of error, we choose a set of modules to exhibit exponential (fast) convergence.
We further provide a toolkit for error analysis and detection, 
which can help a user to identify and mitigate the source of error,
guiding the design of more accurate \Tool programs.

We demonstrate the expressiveness of our language by implementing and simulating common discrete algorithms
such as greatest common divisor, integer division, finding integer square root,
as well as real-valued (analog) algorithms such as computing Euler's number and computing $\pi$.
We implement the \Tool compiler which generates the reactions implementing a high level imperative algorithm, 
and use the \CRNSimulator package~\cite{CRNSimulatorPackage} to manipulate and simulate chemical reactions using Mathematica.
\Tool is an extensible programming language allowing for easy addition of new modules;
we release the open-source version~\cite{CRNPlusPlusGithub} of the tool to enable others make use of it, and extend it further.

\section{Examples}
\label{sec:backgroundAndExample}
In this section we discuss the characteristics of chemical reaction
networks (CRNs) through examples. First, the overall idea of computation in CRNs
is presented, followed by example programs in \Tool. The focus is to give
a high level idea of our technique, while later sections discuss
internal details.

Although historically the focus of the study of CRNs was on understanding the behavior of naturally occurring biological reaction networks, 
recent advancements in DNA synthesis coupled with general methods for realizing arbitrary CRNs with DNA strand displacement cascades~\cite{SoloveichikETAL10DNAUniversalSubstrate} opened the path to engineering
with chemical reactions.  
In this work we are not interested in a way to engineer the
molecules implementing a reaction but
focus on reaction behavior and dynamics. We abstract away molecule
implementation information and denote molecular species with letters
(e.g. $A$).

Molecular systems exhibit complex behaviors governed by chemical reactions.
To give a formal notation of chemical reaction networks, consider the \formulaname~\ref{form:crnExample}~\cite{BuismanETAL09ComputingAlgebraicFunctionsInCRNs}:
\BeginFormula
  \begin{align}
    \label{rxn:mul-one} \ce{$A$ + $B$ &->[1] $A$ + $B$ + $C$} \\
    \ce{$C$ &->[1] $\emptyset$}
  \end{align}
  \caption{Example chemical reaction network}
  \label{form:crnExample}
\EndFormula
\noindent The \formulaname~\ref{form:crnExample} consists of two reactions.
A chemical reaction is defined with \textit{reactants} (left side), \textit{products} (right side),
and \textit{rate constant} which quantifies the rate at which reactants interact to produce products.
To illustrate this, reaction~\ref{rxn:mul-one} is composed of $reactants = \{A,B\}$, $products = \{A,B,C\}$, and rate constant $k=1$.
Since most reactions in \Tool have the rate constant equal to $1$,
from now on we drop the rate constant when writing reactions, unless it is different than $1$.
Note that multiple molecules of same species can be in a list of reactants (analogously for products);
to support this we use the multiset notation.
As an example, to describe reaction: \ce{$A$ + $A$ ->[] $B$} we write $reactants = \{A^2\}$,
where the upper index ($2$) represents multiplicity (number of occurrences). 

It may seem that a molecule of $C$ is produced out of
nothing in reaction~\ref{rxn:mul-one}, since the multiset of reactants
is a submultiset of the products. 
This represents a level of abstraction where \emph{fuel} species that
drive the reaction are abstracted away (i.e., the first reaction
corresponds to \ce{$F$ + $A$ + $B$ ->[] $A$ + $B$ + $C$}). 
Making this assumption allows us to focus on the computationally relevant species.
The choice to use general (non-mass/energy preserving) CRNs is an established convention for DNA strand displacement cascades~\cite{SoloveichikETAL10DNAUniversalSubstrate}.

When the molecular counts of all species are large, and the solution is ``well-mixed'',
the dynamics of the system can be described by ordinary differential equations (mass-action kinetics).
Molecular concentrations are quantified by a system of ODEs,
where the concentration of each species is characterized by the following ODE:  %
\[ \frac{d[S]}{dt}=\sum_{\forall \textit{rxn} \in \textit{CRN}} k(\textit{rxn}) \cdot \textit{netChange}(S,\textit{rxn}) \cdot \prod_{\forall R \in \textit{reactants}(\textit{rxn})} [R]^{m_{\textit{rxn}}(R)}(t) \]
The right side is a sum over reactions in the CRN,
where $k(\textit{rxn})$ is a rate of reaction $\textit{rxn}$,
$\textit{netChange}(S,\textit{rxn})$ is a net change of molecules of $S$ upon triggering of $\textit{rxn}$ (can be negative),
and $m_{\textit{rxn}}(R)$ is the multiplicity of reactant $R$ in reaction $\textit{rxn}$.
To illustrate the general formula, the set of ODEs characterizing \formulaname~\ref{form:crnExample} is:
\[ \frac{d[A]}{dt}=0, \frac{d[B]}{dt}=0, \frac{d[C]}{dt}=[A](t)\cdot[B](t)-[C](t) \]

\begin{wrapfigure}{r}{0.35\textwidth}
  \vspace{-25pt}
  \includegraphics[width=0.35\textwidth]{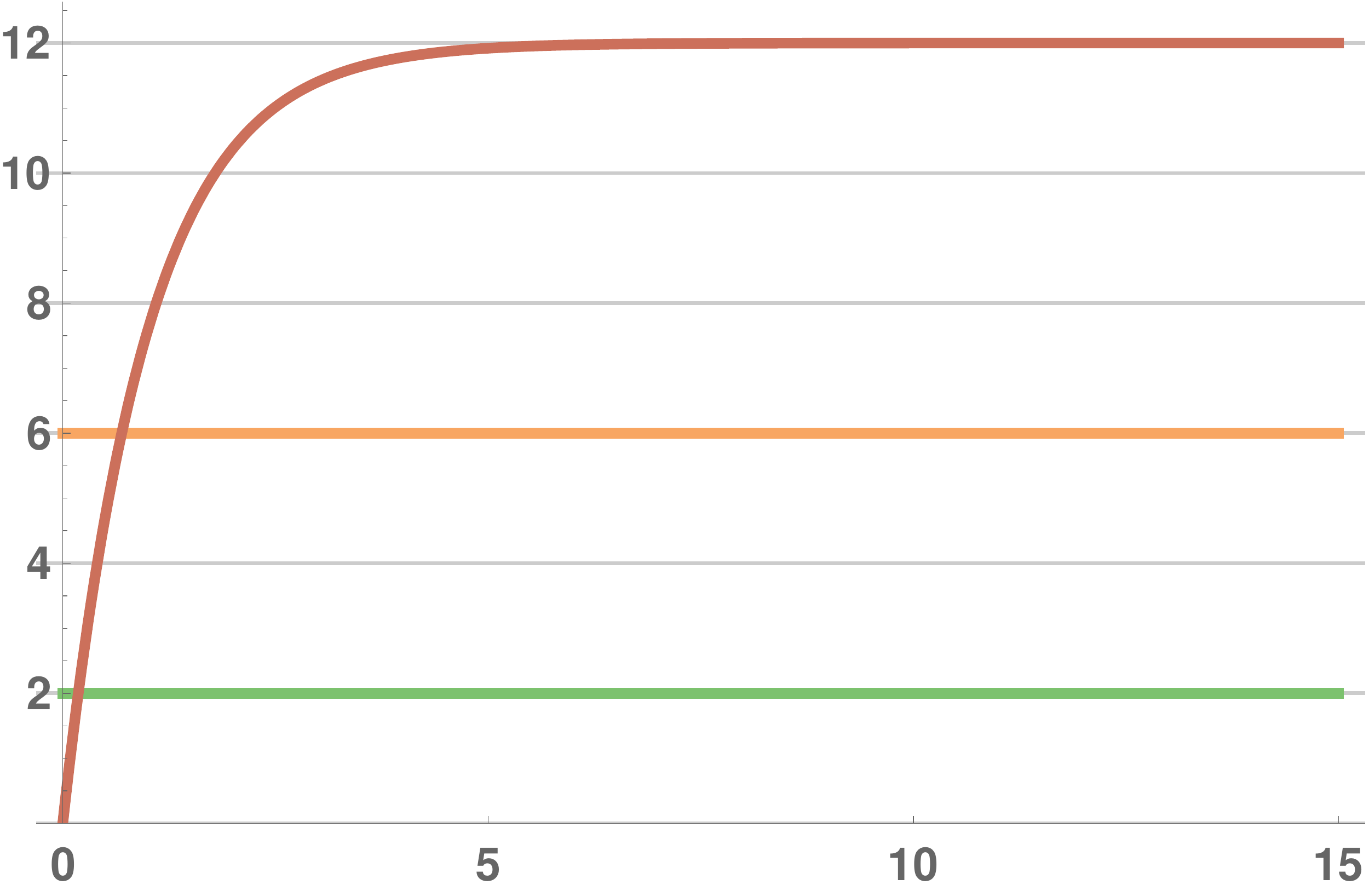}
  \caption{Multiplication CRN. $[A]$ shown in orange, $[B]$ in green, and $[C]$ in red.}
  \label{fig:mul}
  \vspace{-20pt}
\end{wrapfigure}
Since the concentrations $[A]$ and $[B]$ are constant (derivatives zero),
we have $\frac{d[C]}{dt} = [A](0) \cdot [B](0) - [C](t)$.
Thus $[C](t)$ is increasing when smaller than ${[A](0) \cdot [B](0)}$,
decreasing in the opposite case, and does not change when $[C](t) = {[A](0) \cdot [B](0)}$.
Therefore system has a global stable steady-state $[C]={[A](0) \cdot [B](0)}$.
We say that this module computes multiplication,
due to the relation between initial concentrations and concentrations at the steady state. 

We simulate and plot the dynamics of the multiplication CRN, as shown in \figname~\ref{fig:mul}.
Initial concentrations of $A$ and $B$ are $6$ and $2$, respectively,
while the concentration of $C$ approaches value $12$.
Note that the exact value defined by the steady state ($[C](t)=12$) is reached only at the limit of time going to infinity. 
Since the computation has to be done in finite time, the presence of error is unavoidable. 
This error raises challenging issues with programming in chemistry, and
necessitates techniques for controlling it.  One crucial property that
determines the error is the convergence speed of the module. The multiplication
command in \Tool is implemented through the above module, following
the design principles of \emph{convergence speed} and \emph{composability}
described in section~\ref{sec:technique}. Chemical reactions are
abstracted away from a user who can simply write
$\mulModuleNoSpace[a,b,c]$ to multiply.

\Tool is an imperative language, and as such supports sequential execution.
Note that even a simple operation of multiplying and storing into the same variable, e.g.\ $A:=A*B$, requires support for sequential execution.
We use operator ``$:=$'' to relate input and output concentrations; $A:=A*B$ denotes that $[A](t)$ converges to $[A](0)*[B](0)$.
The above implementation of the \mulModule module necessarily assumes that the output species is different from the input species.
Otherwise, $\mulModuleNoSpace[a,b,a]$ goes to infinity or $0$ depending on the value of $B$.
To implement $A:=A*B$, we split the computation into two sequential steps: (1) $C:=A*B$, (2) $A:=C$.
To multiply we use the \mulModule module described above.
For the assignment we use the load module (\ldModule).
To ensure the assignment executes after the multiplication,  %
we catalyze the two modules with the clock species that reach their high values in different phases of the oscillator.
Importantly, the chemical oscillator and clock species are abstracted away from a user,
who simply uses the \stepKeyword construct to order reactions: \stepKeywordNoSpace[\{\mulModuleNoSpace[a,b,c]\}], \stepKeywordNoSpace[\{\ldModuleNoSpace[c,a]\}].

\begin{wrapfigure}{R}{0.35\textwidth}
  \vspace{-20pt}
\begin{algorithmic}[1]
  \Procedure{gcd}{$a,b$}%
  \While{$a\not=b$}
  \If{$a>b$}
  \State $a\gets a - b$
  \Else
  \State $b\gets b - a$
  \EndIf
  \EndWhile\label{euclidendwhile}
  \State \textbf{return} $a$
  \EndProcedure
\end{algorithmic}
   \caption{Euclid's algorithm for computing GCD.}
  \label{fig:euclid}
  \vspace{-20pt}
\end{wrapfigure}

One of the basic blocks of programming languages are conditional branches, executing upon success of a precondition.
Similarly to implementing sequential operations,
we implement conditional execution by activating (through catalysis) some reactions and deactivating others,
depending on a result of condition.
Since no species can be driven to 0 in finite time\footnote{Although certain pathological CRNs can drive concentrations to infinity in finite time (e.g., $2A \rightarrow 3A$), and thereby drive certain other species to $0$ in finite time (e.g., with an additional $B + A \rightarrow A$ ), these cases cannot be implemented with any reasonable chemistry.},
all branches of condition will be active to some extent,
which makes this an interesting source of errors without direct analogy in digital electronics.
Analogous to clock species in sequential execution, reactions are catalyzed by \emph{flag} species to support branching. 
The flag species have high and low values that reflect the result of the comparison.
Our \cmpModule module sets the flag species to reflect the result of the comparison.
In the following example we demonstrate the usage of \cmpModule module and conditional execution.

To demonstrate the expressiveness of our language we showcase
the implementation of Euclid's algorithm (\figname~\ref{fig:euclid}) to compute the greatest common divisor (GCD) of a two numbers. 
The GCD is computed by subtracting the smaller of the values from the larger one until they become equal.

\begin{figure}
  \vspace{-20pt}
  \centering
  \begin{subfigure}[b]{0.45\textwidth}
\begin{lstlisting}[
  language=Mathematica,
  %
  %
  %
  escapechar=\#
]
crn = {
  conc[a,a0], 
  conc[b,b0],
  step[{
    ld[a, atmp],
    ld[b, btmp], 
    cmp[a,b] 
  }],
  step[{
    ifGT[{ sub[atmp,btmp,a] }],
    ifLT[{ sub[btmp,atmp,b] }]
  }]
};
\end{lstlisting}
     \caption{GCD implementation}
    \label{fig:gcdLst}
  \end{subfigure}
  \hfill
  \begin{subfigure}[b]{0.47\textwidth}
    \includegraphics[width=\textwidth]{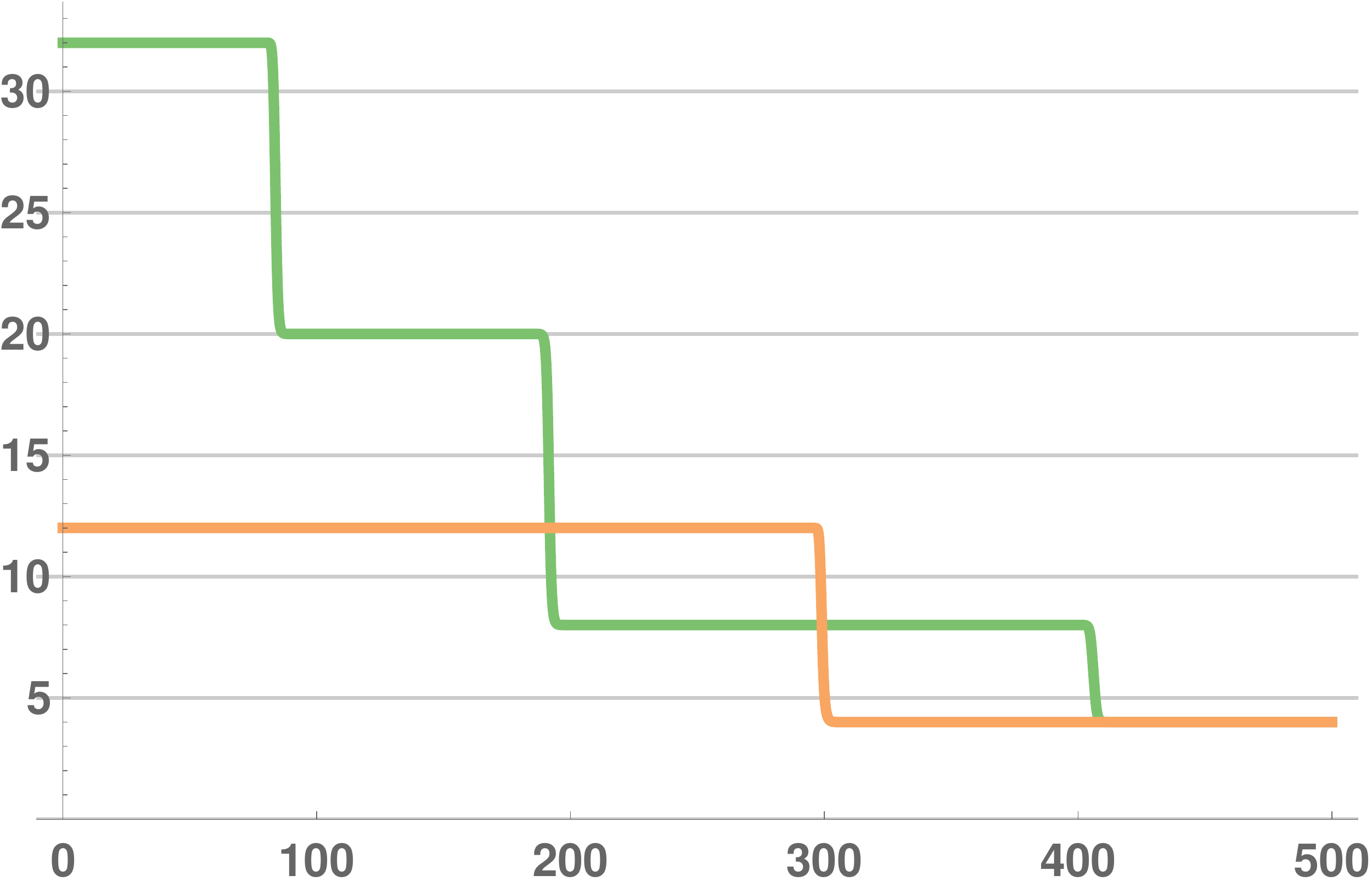}
    \caption{Dynamic simulation of the GCD program for \mbox{$a0=32$}, \mbox{$b0=12$}.
      Concentrations of $a$ (green), and $b$ (orange) are shown in function of time.}
    \label{fig:gcdSimulation}
  \end{subfigure}
  \caption{Implementation of Euclid's algorithm for computing GCD in
    \Tool (left), simulation results of the implementation (right).}
  \label{fig:gcd}
\end{figure}

\figname~\ref{fig:gcdLst} shows the implementation of Euclid's algorithm in \Tool.
Lines 2-3 define the initial concentrations of species $a$ and $b$,
where constants $a0$ and $b0$ are values for which GCD is computed.
To order the execution, the \stepKeyword construct is used.
Multiple instructions that do not conflict with each other can be part of the same step and they are executed in parallel.
In the first step $a$ and $b$ are stored into temporary variables
and compared, setting the flag species to reflect the result of the comparison.
The second step uses the result of the previous comparison, and effectively stores $a-b$ into $a$ if $a>b$, and vice versa.
Since the same species cannot be used as both input and output to \subModule module, temporary variables are used ($atmp$ and $btmp$).
Steps repeatedly execute due to the oscillatory behavior of the clock species, thus implementing looping behavior by default;
the steps can be viewed as being inside of the `forever' loop.
\Tool, in addition to the language and compiler to chemical reactions, is connected to the simulation backend that enables convenient testing for correctness.

We show a simulation of the GCD program in \figname~\ref{fig:gcdSimulation} where GCD(32,12) is computed.
Although not visible in the plot we can identify a number of non-idealities in the \Tool implementation.
First, modules converge to correct values only in limit of time going to infinity; thus for example the first subtraction (second step) will set $a$ to value close to but not exactly $20$.
Second, modules that should not be executing cannot be completely turned off, and so when the first step executes the second does so as well (of course, in much smaller extent). 
Analogously the two subtraction operations in the second step are supposed to be mutually exclusive, and yet they co-occur to some extend.
More discussion on the sources of error is provided in section~\ref{sec:technique:errorEval}.

\noindent
\vspace{-20pt}

In addition, we implement a set of algorithms in
(a) discrete space---counter, factorial, integer division, integer square root, as well as in
(b) continuous space---by implementing \Tool programs that approximate value of \textit{Euler's} constant and $\pi$.
These examples are shown in section~\ref{sec:evaluation}.

\section{Technique}
\label{sec:technique}
This section explains \Tool, both the underlying constructs used to
build it, as well as high level primitives that represent the language
itself. We start by presenting high-level modules that are at the
core of \Tool (section~\ref{sec:technique:modules}), followed by
explanation of how the sequential behavior is achieved
(section~\ref{sec:technique:oscillator}), after which we give an overview of \Tool
grammar (section~\ref{sec:technique:grammar}), and finally we
discuss the error detection and analysis tools we provide
(section~\ref{sec:technique:errorEval}).

\subsection{Modules}
\label{sec:technique:modules}
Modules represent the core of \Tool,
and in their form are somewhat analogous to the instruction set architecture (ISA) in machine languages.
Modules implement basic operations such as load, add, subtract, multiply, compare.
We provide the exhaustive list of modules in Table~\ref{table:modules}.
Importantly, \Tool is extensible, and supports easy addition of new modules.

There are multiple ways of computing addition and other operations in
chemistry. As mentioned in the previous section, our implementation choice is led by two basic principles:
(a) convergence speed, and (b) composability.

\Subsubsection{Convergence speed}
\label{sec:convergenceSpeed}

To provide intuition about the convergence speed of modules we analyze the following CRN:

\BeginFormula
  \begin{align}
    \ce{$A$ + $B$ ->[] $\emptyset$} \nonumber
  \end{align}
  \caption{Simple CRN.}
  \label{form:crnSimple}
\EndFormula

The ODEs describing the above CRN are: $\frac{d[A]}{dt}=-[A](t)*[B](t)$ and $\frac{d[B]}{dt}=-[A](t)*[B](t)$.
Since the amount of $B$ decreases with the same speed as $A$, we can express the dynamics of the system in terms of $D_0=[B](0)-[A](0)$:
\[ \frac{d[A]}{dt}=-[A](t)*([A](t)+D_0) \]
If $D_0 \neq 0$, the solution is: \[[A](t)=\frac{[A](0) D_0}{-[A](0) + [A](0) e^{D_0t} + D_0 e^{D_0t}} \]
To consider the convergence speed we look at the non-constant part of the equation.
Due to the factor $e^{-t}$ the decrease of the non-constant part is exponential, and we say that the convergence speed is \textit{exponential}.
If $D_0 > 0$ ($[B](0)>[A](0)$) terms with exponential factors tend to infinity, and $[A]$ to zero.
Conversely, when $D_0<0$, exponential factors converge to zero, and $[A]$ to $-D_0$.

When $D_0 \neq 0$ so far; by solving the ODE for $D_0=0$ we get a following:
\[ [A](t)=\frac{[A](0)}{1 + [A](0) t} \]
In this equation, non-constant part is not anymore exponential, but linear; thus we say that the convergence speed is \textit{linear}.
To summarize, \formulaname~\ref{form:crnSimple} exhibits exponential convergence speed when $[A](0) \neq [B](0)$, and linear otherwise.

\Tool models computation in which an instruction executes in a step (clock phase), and computation in subsequent steps depends on values computed in previous steps.
In this reason, it is of a crucial importance to achieve a high precision of computation until the end of a current step;
and to achieve this goal it is necessary to ensure a high convergence speed.
For this reason we want to ensure that all of our modules exhibit exponential convergence. 
So, for example, we do not want to use the reaction $A + A \to C$ to compute division by 2 since it exhibits only linear convergence (corresponding to the $A=B$ analysis above).

\Subsubsection{Composability}

To explain composability we analyze two different CRNs that perform addition (i.e., compute $C:=A + B$).
\BeginFormula
  \begin{align}
    \ce{$A$ &->[] $A$ + $C$} \nonumber \\
    \ce{$B$ &->[] $B$ + $C$} \nonumber \\
    \ce{$C$ &->[] $\emptyset$} \nonumber
  \end{align}
  \caption{Addition CRN (preserves inputs). Inputs: $A$ and $B$, output: $C$.}
  \label{form:crnAdd}
\EndFormula
\BeginFormula
  \begin{align}
    \ce{$A$ &->[] $C$} \nonumber \\
    \ce{$B$ &->[] $C$} \nonumber
  \end{align}
  \caption{Addition CRN (destroys inputs). Inputs: $A$ and $B$,
    output: $C$.}
  \label{form:crnAddDestructive}
\EndFormula
Although both of these CRNs compute addition (this is evident for \formulaname~\ref{form:crnAddDestructive}; more detailed derivation is presented for \formulaname~\ref{form:crnAdd} in section~\ref{sec:addmodule}), they behave very differently when composed with other modules.
For example, to compute $W:=(X*Y)+Z$, we can combine the \mulModule module (\formulaname~\ref{form:crnExample}) with $X$ and $Y$ as inputs and some species $K$ as output, with an addition module with $K$ and $Z$ as inputs and $W$ as output.
If we use \formulaname~\ref{form:crnAdd} for addition, note that $K$ is not changed by it (the input acts catalytically), and thus $K$ approaches $X * Y$ as expected, allowing the addition module to correctly compute $(X*Y)+Z$ in the limit.
In contrast, if  we use \formulaname~\ref{form:crnAddDestructive} to perform addition, then by consuming $K$ the addition module will affect the equilibrium of the multiplication module, driving it lower than the desired value $X*Y$. 
In this way \formulaname~\ref{form:crnAddDestructive} is not composable, while \formulaname~\ref{form:crnAdd} allows for the correct composition.

More generally, for any set of modules that: (1) use inputs catalytically (does not produce or consume inputs), (2) exhibit exponential convergence, and (3) have a unique stable steady state, a CRN composed of such modules also has the above three properties. A proof can be found in Buisman et al.~\cite{BuismanETAL09ComputingAlgebraicFunctionsInCRNs}. From this it follows that the composed CRN of the \mulModule module and \formulaname~\ref{form:crnAdd} exhibits the exponential convergence speed, and has a stable steady state defined by $W:=(X*Y)+Z$.

Note that the above discussion of composability concerns reactions occurring within a single step construct (i.e., we discuss correct computations when reactions occur concurrently).
Even non-composable reactions can be composed by separating them in time via the step construct. 
However, in order to keep the number of steps low and ensure faster computation we aim to make reactions composable when possible.
With this, we have set up the two main design criteria (convergence speed and composability) for the modules, and we next describe the \Tool modules.

\Subsubsection{\ldModuleCapitalized Module}
Loads the value from source (first argument) into a destination
(second argument). The CRN used for load operation is following:
\BeginFormula
  \begin{align}
    \ce{$A$ &->[] $A$ + $B$} \nonumber \\
    \ce{$B$ &->[] $\emptyset$} \nonumber
  \end{align}
  \caption{Load CRN}
  \label{form:crnLd}
\EndFormula
$A$ is the input and $B$ is the output species. This
module, similar to \addModule (see next section), has exponential convergence speed~\cite{BuismanETAL09ComputingAlgebraicFunctionsInCRNs}. In addition,
the concentration of input species is constant, thus ensuring composability.

\Subsubsection{\addModuleCapitalized Module}  \label{sec:addmodule}
Adds two values (first and second argument) and stores the result into destination (third argument).
The Add CRN is shown in \formulaname~\ref{form:crnAdd}.

By solving the ODEs that characterize $[C](t)$ we get the following equation:
$ [C](t)=[A](0) + [B](0) + ([C](0) - [A](0) - [B](0)) \cdot \euler^{-t} $.
From the equation it follows that $[C]$ converges to $[A](0) + [B](0)$, and thus we say the CRN performs addition.
Moreover, the CRN exhibits exponential convergence.

\Subsubsection{\subModuleCapitalized Module}
Subtracts the second input value from the first and stores into the destination (third argument).
\BeginFormula
  \begin{align}
    \ce{$A$ &->[] $A$ + $C$} \nonumber \\
    \ce{$B$ &->[] $B$ + $H$} \nonumber \\
    \ce{$C$ &->[] $\emptyset$} \nonumber \\
    \ce{$C$ + $H$ &->[] $\emptyset$} \nonumber
  \end{align}
  \caption{Subtraction CRN}
  \label{form:crnSub}
\EndFormula
The above CRN was generated via evolutionary algorithms~\cite{BuismanETAL09ComputingAlgebraicFunctionsInCRNs};
by analyzing its system of ODEs, the network computes truncated subtraction:
\begin{equation}
  [C]=
    \begin{cases}
      [A]-[B], & \text{if}\ [A]>[B] \\
      0, & \text{otherwise}
    \end{cases}
\end{equation}
Input species $A$ and $B$ are not affected and the property of composability is satisfied.
Neither we nor Buisman et al.\ found the analytical solution;
however, our simulation results show that the module converges exponentially quickly unless $A=B$ (see the section~\ref{sec:convergenceSpeed} for an analogous, easy to analyze case).
In a case inputs, $A$ and $B$, are close to each other the computation error is higher.
The error evaluation tools (section~\ref{sec:technique:errorEval}) help in detecting and analyzing problematic cases (e.g., where $A$ and $B$ are close),
thus enabling a user to redesign the CRN.
In our examples, $A$ and $B$ usually differ by at least $1$.
Runtime assertions in the simulation package that automatically notify a user about these kind of problems would help identify the source of the error.
Note that many algorithms can be refactored to reduce the error (see section~\ref{sec:discAndConcl}).

\Subsubsection{\mulModuleCapitalized Module}
Multiplies inputs (first and second argument) and stores into
destination (third argument). The multiplication CRN is shown in
section~\ref{sec:backgroundAndExample}. This CRN does not affect
inputs and has exponential convergence speed~\cite{BuismanETAL09ComputingAlgebraicFunctionsInCRNs}.

\vspace{5pt}
We have presented modules for performing arithmetic operations (\ldModule, \addModule, \subModule, \mulModule).
These modules are implemented within a single \emph{step}.
Multiple modules can be executed in parallel within a single \emph{step} as long as there is no cyclic dependence between species: for example \mulModuleNoSpace[a,b,c] and \addModuleNoSpace[c,d,a] forms a cycle,
the output of the \mulModule is input to the \addModule, and vice versa.
Also, the CRN implementation imposes the restriction that same species cannot be used as both input and output to the same module (which is really a cycle of length 1).
We now introduce the \cmpModule module providing for conditional execution, which is executed in two \emph{steps}.

\Subsubsection{\cmpModuleCapitalized Module}
\label{sec:cmpModule}
Compares the two values, and produces signals (flag species) informing
which value is greater or if they are equal.

The \cmpModule module is implemented using two sequentially executed sets of reactions,
which trigger in consecutive clock phases.
In the first phase, the inputs ($X$ and $Y$) are mapped to flag species \xgty and \xlty.
Values are mapped to the range [0--1], by setting the initial concentrations of \xgty and \xlty such that their sum is $1$. %
If, for example, $[X]=80$ and $[Y]=20$, flag species \xgty and \xlty converge to $0.8$ and $0.2$, respectively.
\footnote{This convergence happens irrespective of the initial concentrations of the flag species \xgty and \xlty (as long as they sum to $1$), so they do not have to be reset before the mapping.}
The mapping is done in order to preserve the original values of the inputs ($X$ and $Y$), considering that the next phase of comparison consumes the compared values (flags).
The mapping CRN is shown in \formulaname~\ref{form:crnNormalize},
and exhibits exponential convergence speed according to our analysis.
\BeginFormula
  \begin{align}
    \ce{$X_{gtY}$ + $Y$ &-> $X_{ltY}$ + $Y$} \nonumber \\
    \ce{$X_{ltY}$ + $X$ &-> $X_{gtY}$ + $X$} \nonumber
  \end{align}
  \caption{CRN for mapping compared values}
  \label{form:crnNormalize}
\EndFormula
The goal of the second phase of comparison is to detect which value is greater.
We use a chemical \textit{Approximate Majority} (AM) algorithm~\cite{CardelliAttila12CellCycleSwitch}
to detect if \xgty or \xlty is in the majority.
All molecules of the less populous species convert to the more populous species.
The AM reactions are:
\BeginFormula
  \begin{align}
    \ce{$X_{gtY}$ + $X_{ltY}$ &->[] $X_{ltY}$ + $B$} \nonumber \\
    \ce{$B$ + $X_{ltY}$ &->[] $X_{ltY}$ + $X_{ltY}$} \nonumber \\
    \ce{$X_{ltY}$ + $X_{gtY}$ &->[] $X_{gtY}$ + $B$} \nonumber \\
    \ce{$B$ + $X_{gtY}$ &->[] $X_{gtY}$ + $X_{gtY}$} \nonumber
  \end{align}
  \caption{Approximate Majority CRN}
  \label{form:crnAM}
\EndFormula
Recall that the normalization CRN is run in a previous step, so:
\begin{equation}
  \label{eq:normalizedFlags}[X_{gtY}](0) + [X_{ltY}](0) = 1
\end{equation}
Then in the AM network, the species (\xgty, \xlty, $B$) converge to values (1, 0, 0) if $X_{gtY}(0) > X_{ltY}(0)$ and (0, 1, 0) if $X_{gtY}(0) < X_{ltY}(0)$.
In the subsequent step constructs, the species \xgty are used as a catalysts in reactions that execute when \xgtyEq, and the species \xlty for the opposite case.

We now argue that CRN~\ref{form:crnAM} exhibits exponential convergence.
Due to normalization,~\ref{eq:normalizedFlags} holds, and moreover \mbox{[\xgtyNoSpace](t) + [\xltyNoSpace](t) + [B](t) = 1} (for all time t).
Taking the previous equality into account the ODEs of AM are:
\begin{align}
  \label{eq:dXgtY} \frac{d[X_{gtY}]}{dt} &= [X_{gtY}](t)(1-[X_{gtY}](t)-2[X_{ltY}](t)) \\
  \label{eq:dXltY} \frac{d[X_{ltY}]}{dt} &= [X_{ltY}](t)(1-[X_{ltY}](t)-2[X_{gtY}](t))
\end{align}
Consider the function $f(t) = \frac{(X_{gtY}-X_{ltY})^3}{X_{gtY}X_{ltY}}$.
(The choice of this function is guided by the closed form solution for $t$ as a function of $X_{gtY}(t)$ and $X_{ltY}(t)$~\cite{PerronETAL09BinaryConsensus}.)
Taking the derivative of $f(t)$ with respect to $t$ and substituting~\ref{eq:dXgtY} and~\ref{eq:dXltY} results in $f(t)$ again.
This means that $f(t)$ has a solution $f(t) = Ce^t$, which is unique by the Picard-Lindel\"{o}f theorem.
Therefore, $Ce^t = \frac{(X_{gtY}-X_{ltY})^3}{X_{gtY}X_{ltY}} < 1/(X_{gtY}X_{ltY})$ since the concentrations are at most $1$.
This constrains $X_{gtY}X_{ltY} < C^{-1} e^{-t}$, which implies that $X_{gtY}$ or $X_{ltY}$ converges to $0$ exponentially quickly.

If input species are initially equal ([\xgtyNoSpace](0) = [\xltyNoSpace](0)) then the system preserves their equality,
and following ODE holds:
$[X_{gtY}]'(t) = 2[X_{gtY}](t)(1 - 3[X_{gtY}](t))$.
This is a logistic differential equation which converges exponentially quickly to values (1/3, 1/3, 1/3).)
Even though convergence is exponential we do not know how the time changes as a function of the difference between the input species, when the difference is small.
Thus, even if the compared species are close to each other, the system may exhibit undesirably slow convergence to (1, 0, 0) or (0, 1, 0);
As we later explain, our error evaluation framework can help detect such cases.

\textbf{Equality checking.}  Due to the ever-present error
in chemical computation, checking for equality is actually
approximate-equality checking. Consider having a chemical program with
real values, then if the values are close to each other it is
impossible to tell if they are actually equal but affected with
error, or they represent different real valued signals. Due to this
issue, while comparing for equality is impossible, we compare for
$\epsilon$-range equality.
For discrete algorithms we use equality checking with $\epsilon=0.5$,
allowing easy comparison of the integer values (e.g., values in range $(2.5,3.5)$ are considered to be equal to $3$).

To support equality checking we compare $x + \epsilon$ with $y$ (generating
signals \xgty and \xlty), and at the same time compare $y + \epsilon$ with
x (generating signals \ygtx and \yltx). Combining the signals of
the two comparisons gives the desired result: If \xeqyEq, signal \xgty is
high (\xlty low) and \ygtx is high (\yltx low) due to the added
offset. To execute a reaction upon equality both \xgty and \ygtx are
used catalytically. If \xgtyEq, signal \xgty is high (\xlty low) and
\yltx is high (\ygtx low), so both \xgty and \yltx should be used
catalytically. Symmetrically for \xltyEq, both \xlty and \ygtx are
used catalytically. 
After calling \cmpModule in a step, a user can use instructions
\gtKeyword (\textit{greater than}), \geKeyword (\textit{greater or equal}), \eqKeyword (\textit{equal}), \ltKeyword (\textit{less than}), \leKeyword (\textit{less or equal})
in subsequent steps to conditionally execute reactions.
Note that the flags are active until the next call to the \cmpModule module.

\begin{table}[t!] %
  \scriptsize
  \centering
  \setlength{\tabcolsep}{8pt}
  \renewcommand{\arraystretch}{1.5}
  \begin{tabularx}{\textwidth}{ | c | c | X | X |}
    \hline
    Type   & Restrictions & Output (Steady State) & CRN \\
    \hline
    \ldModuleNoSpace [A,B]  & $B \not\equiv A$                & $B := A$ &
    \parbox[c]{1pt}{
      \vspace{-5pt}
      \begin{align}
        \ce{$A$ &->[] $A$ + $B$} \nonumber \\
        \ce{$B$ &->[] $\emptyset$} \nonumber
      \end{align}
      \vspace{-10pt}
    } \\
    \hline
    \addModuleNoSpace [A,B,C] & $C \not\equiv A \land C \neq B$ & $C := A + B$ &
    \parbox[c]{1pt}{
      \vspace{-5pt}
      \begin{align}
        \ce{$A$ &->[] $A$ + $C$} \nonumber \\
        \ce{$B$ &->[] $B$ + $C$} \nonumber \\
        \ce{$C$ &->[] $\emptyset$} \nonumber
      \end{align}
      \vspace{-10pt}
    } \\
    \hline
    \subModuleNoSpace [A,B,C] & $C \not\equiv A \land C \not\equiv B$ &
    \parbox[c]{1pt}{
      \[
      C:=
      \begin{cases}
        \tiny
        A - B, &\text{$A > B$} \\
        0, &\text{otherwise}
      \end{cases}
      \]
    } &
    \parbox[c]{1pt}{
      \vspace{-5pt}
      \begin{align}
        \ce{$A$ &->[] $A$ + $C$} \nonumber \\
        \ce{$B$ &->[] $B$ + $H$} \nonumber \\
        \ce{$C$ &->[] $\emptyset$} \nonumber \\
        \ce{$C$ + $H$ &->[] $\emptyset$} \nonumber
      \end{align}
      \vspace{-10pt}
    } \\
    \hline
    \mulModuleNoSpace [A,B,C] & $C \not\equiv A \land C \not\equiv B$ & $C := A \cdot B$ &
    \parbox[c]{1pt}{
      \vspace{-5pt}
      \begin{align}
        \ce{$A$ + $B$ &->[] $A$ + $B$ + $C$} \nonumber \\
        \ce{$C$ &->[] $\emptyset$} \nonumber
      \end{align}
      \vspace{-10pt}
    } \\
    \hline
    \divModuleNoSpace [A,B,C] & $C \not\equiv A \land C \not\equiv B$ & $C := A / B$ &
    \parbox[c]{1pt}{
      \vspace{-5pt}
      \begin{align}
        \ce{$A$ &->[] $A$ + $C$} \nonumber \\
        \ce{$B$ + $C$ &->[] $B$} \nonumber
      \end{align}
      \vspace{-10pt}
    } \\
    \hline
    \sqrModuleNoSpace [A,B] & $B \not\equiv A$                & $B := \sqrt{A} $ &
    \parbox[c]{1pt}{
      \vspace{-5pt}
      \begin{align}
        \ce{$A$ &->[1] $A$ + $B$} \nonumber \\
        \ce{$B$ + $B$ &->[\frac{1}{2}] $\emptyset$} \nonumber
      \end{align}
      \vspace{-10pt}
    } \\
    \hline
    \cmpModuleNoSpace [A,B] & $A \not\equiv B$                & Sets flag species & * Two CRNs (mapping and AM) triggering in a two consecutive phases (as discussed in section~\ref{sec:cmpModule}) \\
    \hline
  \end{tabularx}
  \vspace{10pt} %
  \caption{\Tool Modules.
    The first column denotes the type of the module.
    The restrictions column imposes compile-time restrictions for using modules,
    here $\not\equiv$ is used to mean different species (not values).
    The output column shows the value of outputs at the steady state.
    Finally, the CRN column shows chemical reactions implementing the module.}
  \label{table:modules}
  \vspace{-20pt}
\end{table}

\subsection{Sequential Execution}
\label{sec:technique:oscillator}

\begin{figure}[!t]
  \centering
  \includegraphics[width=0.4\textwidth]{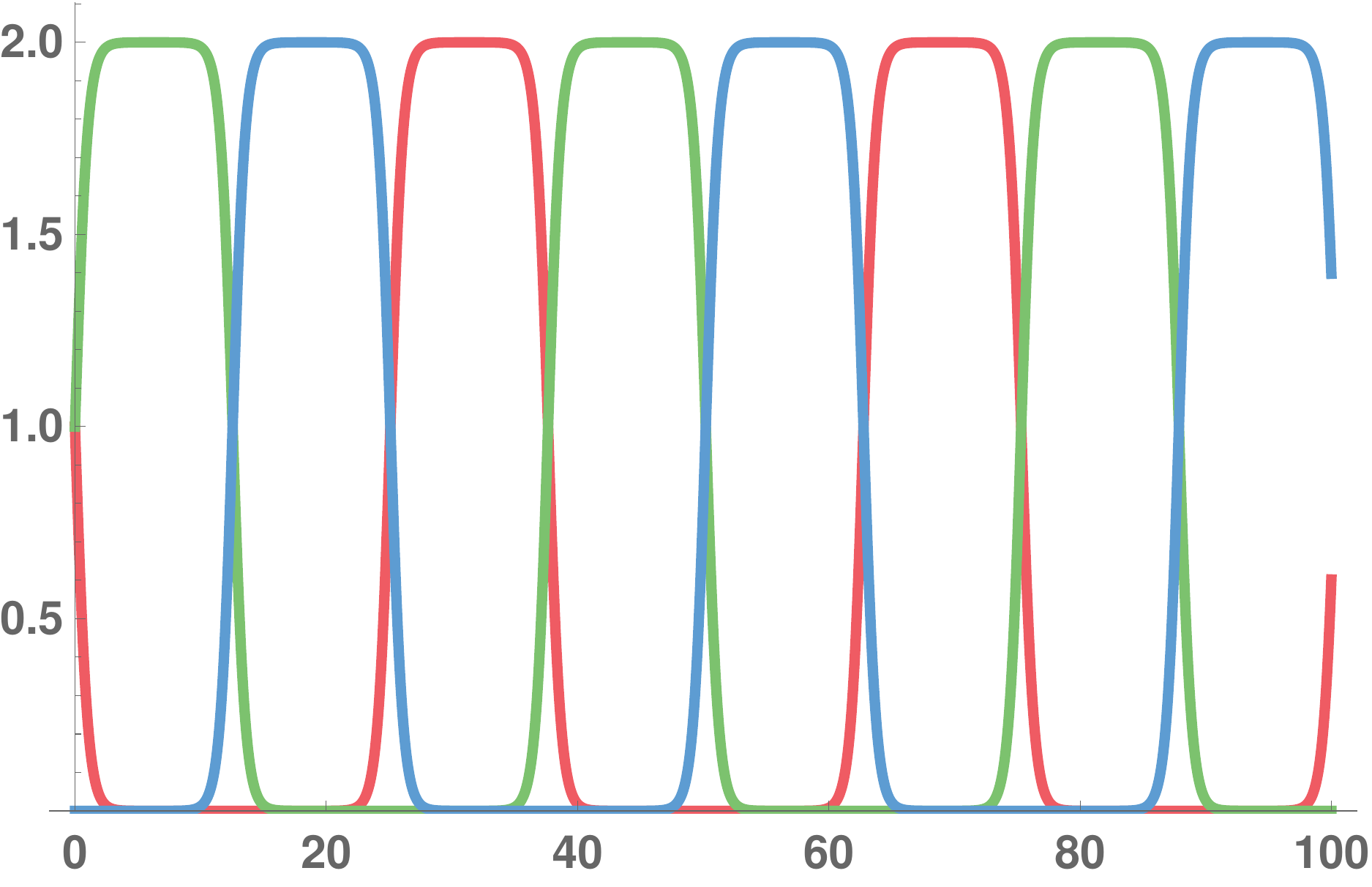}
  \caption[]{Chemical oscillator containing 3 species: $X_1$ (red), $X_2$ (green), and $X_3$ (blue).}
  \label{fig:oscillator}
  \vspace{-10pt}
\end{figure}

\Tool allows programming in a sequential manner, despite the intrinsically parallel nature of CRNs.
To model sequential execution it is necessary to isolate two reactions from co-occurring, and control the order in which they happen.
The key construct we rely on to achieve these goals is a chemical oscillator.

A chemical oscillator is a CRN in which the concentrations of species oscillate between low and high values.
The oscillatory CRN~\cite{Lachmann95ComputationallyCompleteAntColony} we use is described with a following set of reactions:
\BeginFormula
  \begin{align}
    i=1,...,n-1: \ce{$X_i$ + $X_{i+1}$ &-> 2$X_{i+1}$} \nonumber \\
    \ce{$X_n$ + $X_1$ &-> 2$X_1$} \nonumber
  \end{align}
  \caption{Oscillator CRN}
  \label{form:oscillator}
\EndFormula

Concentrations of the clock species ($X_i$) oscillate (see~\figname~\ref{fig:oscillator}).
Different clock species have different oscillation phase and reach minimum and maximum at different times.
To control the rate at which a reaction fires, clock species are added as both reactant and product (catalyst),
in that way preventing reactions from co-occurring and ordering them (see \formulaname~\ref{form:orderingReactions}).
While overlap between the clock species exists, it is small and thus enables sequential execution.
To ensure the small overlap, in \Tool we use every third clock species, i.e. $X_3$, $X_6$, $X_9$, etc., to catalyze the reactions that execute at different time moments.

\BeginFormula
  \begin{align}
    \ce{$A$ -> $B$} && \ce{$A$ + $X_3$ -> $B$ + $X_3$} \nonumber \\
    \ce{$B$ -> $C$} && \ce{$B$ + $X_6$ -> $C$ + $X_6$} \nonumber
  \end{align}
  \caption{Ordering reactions: original reactions (left), ordered (right).}
  \label{form:orderingReactions}
\EndFormula

The chemical oscillator is abstracted from a \Tool user,
who can order reactions using the \stepKeyword construct.
Reactions in different \steps are isolated from each other through clock species acting catalytically.

Non-conflicting instructions can be part of the same step.
Splitting instructions across multiple steps is needed in a case of
a) conditional execution---comparison needs to be done before conditional execution is possible;
b) reading and writing to the same species---this is not possible within the step (as discussed earlier), and requires temporal ordering.
The number of clock species used is determined by the number of \texttt{step} instructions.
Each step requires three clock species,
with the exception of steps in which \cmpModule module is used,
for which six clock species are used.
The \cmpModule module requires two temporarily ordered operations: normalization and approximate majority, and thus six clock species are used.
The oscillatory behavior of the clock species causes steps to get repeated eventually,
causing the loop-like behavior.

\subsection{Grammar}
\label{sec:technique:grammar}

\begin{figure}[!t]
  \centering
  \begin{minipage}{.5\textwidth}
\begin{BNF}
  \footnotesize
  \begin{grammar}
    <\crnNonterminal> ::= `\crnKeywordNoSpace = {'<\rootStatementListNonterminal>`}'

    <\rootStatementListNonterminal> ::= <\rootStatementNonterminal>
    \alt <\rootStatementNonterminal> `,' <\rootStatementListNonterminal>

    <\rootStatementNonterminal> ::= <\concStatementNonterminal>
    \alt <\stepStatementNonterminal>

    <\concStatementNonterminal> ::= `conc['<species>`,'<number>`]'

    <\stepStatementNonterminal> ::= `\stepKeywordNoSpace['\nestedStatementListNonterminal`]'

    <\nestedStatementListNonterminal> ::= <\nestedStatementNonterminal>
    \alt <\nestedStatementNonterminal> `,' <\nestedStatementListNonterminal>

    <\nestedStatementNonterminal> ::= <\rxnStatementNonterminal>
    \alt <\arithmeticStatementNonterminal>
    \alt <\cmpStatementNonterminal>
    \alt <\conditionalStatementNonterminal>

    <\rxnStatementNonterminal> ::= `rxn['<\expressionNonterminal>`,'<\expressionNonterminal>`,'<number>`]'

    <\moduleStatementNonterminal> ::= `\ldModuleNoSpace['<species>`,'<species>`]'
    \alt `\addModuleNoSpace['<species>`,'<species>`,'<species>`]'
    \alt `\subModuleNoSpace['<species>`,'<species>`,'<species>`]'
    \alt `\mulModuleNoSpace['<species>`,'<species>`,'<species>`]'
    \alt `\divModuleNoSpace['<species>`,'<species>`,'<species>`]'
    \alt `\sqrModuleNoSpace['<species>`,'<species>`]'
    \alt `\cmpModuleNoSpace['<species>`,'<species>`]'

    <\conditionalStatementNonterminal> ::= `\gtKeywordNoSpace['<\nestedStatementListNonterminal>`]'
    \alt `\geKeywordNoSpace['<\nestedStatementListNonterminal>`]'
    \alt `\eqKeywordNoSpace['<\nestedStatementListNonterminal>`]'
    \alt `\ltKeywordNoSpace['<\nestedStatementListNonterminal>`]'
    \alt `\leKeywordNoSpace['<\nestedStatementListNonterminal>`]'

    <\expressionNonterminal> ::= <species> \{ `+' <species> \}
  \end{grammar}
  \vspace{-10pt}
  \caption{\Tool Grammar}
  \label{lst:grammar}
\end{BNF}
   \end{minipage}
\end{figure}

\Tool grammar expresses syntactic rules of \Tool programs, and is shown in Listing~\ref{lst:grammar}.
A program consists of a sequence of \rootStatementNonterminal\ statements, where \rootStatementNonterminal\ can be either \concStatementNonterminal\ which defines the initial concentration of species, or \stepStatementNonterminal\ which orders execution.
Furthermore, a \stepStatementNonterminal\ statement is divided into a list of \nestedStatementNonterminal\ statements, where each \nestedStatementNonterminal\ is either: (1) \rxnStatementNonterminal\ which explicitly defines a reaction, (2) \moduleStatementNonterminal\ which defines a module, or (3) \conditionalStatementNonterminal\ which conditinally executes a block based on the result of a previous comparison.
Based on the result of the comparison, only the appropriate conditional blocks are executed: \gtKeyword (\textit{greater than}), \geKeyword (\textit{greater or equal}), \eqKeyword (\textit{equal}), \ltKeyword (\textit{less than}), \leKeyword (\textit{less or equal}).
Note that the semantic rules of \Tool require that comparison is done in a step prior to the conditional execution.

The grammar can be easily extended; e.g., new modules can be added to the list of \moduleStatementNonterminal\ nonterminals.

\subsection{Error Evaluation}
\label{sec:technique:errorEval}

\begin{figure}[!t]
  \centering
  \includegraphics[width=0.45\textwidth]{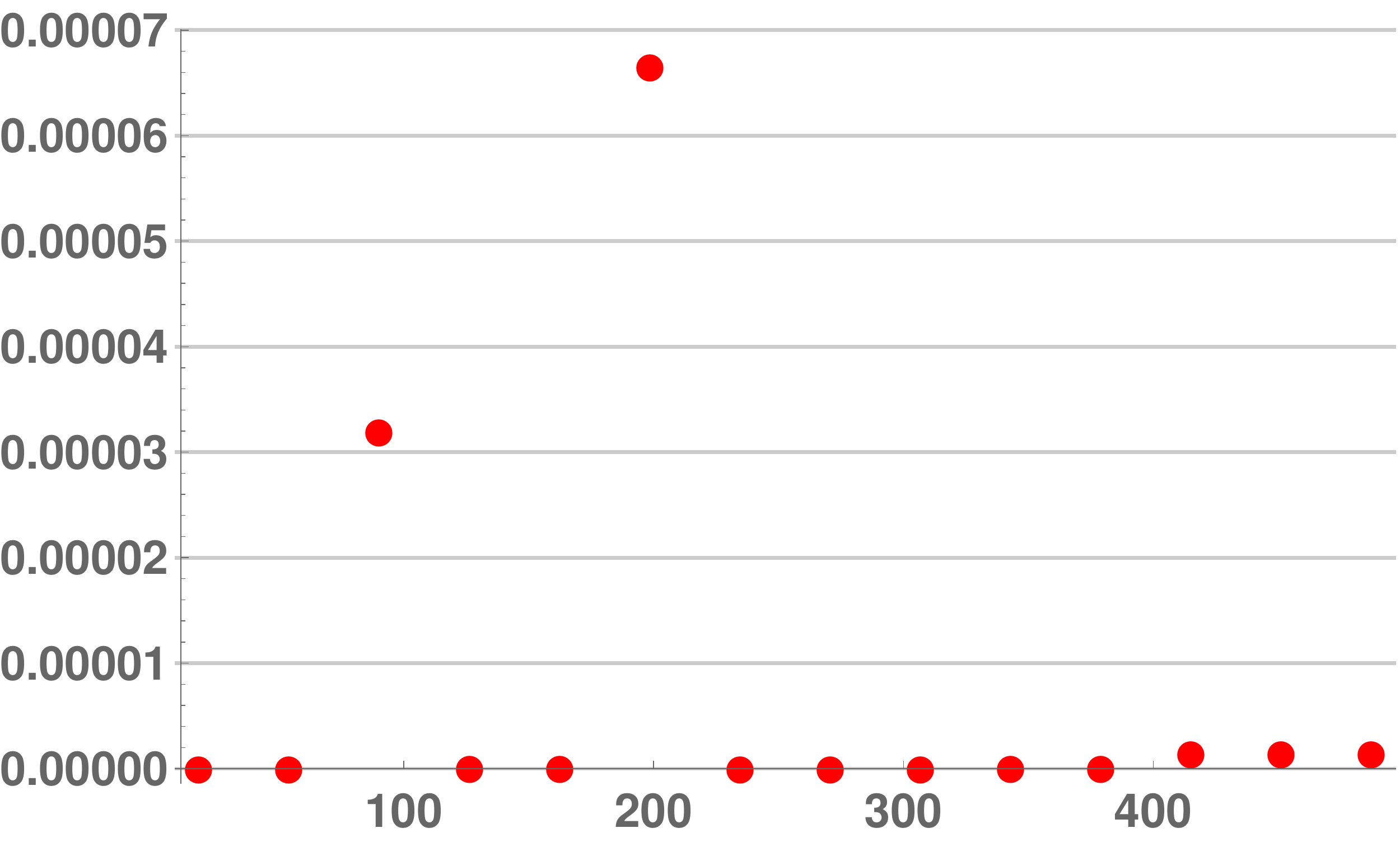}
  \caption[]{Error evaluation of species $a$ from GCD program.}
  \label{fig:gcdError}
  \vspace{-10pt}
\end{figure}

Programming chemistry is inherently error-prone.
We identify three specific sources of error in \Tool.
First, CRNs converge asymptotically---only in the limit is the correct value reached---thus leaving certain amount of error in a finite time.
Second, we cannot completely turn off modules which are not supposed to be currently executing, whether they belong to another sequential step, or to another branch of execution.
Third, comparison has to take into account possible error in the compared values.

Our design decisions were based on minimizing the error; however since error cannot be avoided altogether, we provide a toolkit that helps in error analysis and guiding the CRN (program) design.
Using the tool, users can, for any species of interest, track the difference between the correct value, and the (simulated) value in chemistry.
For example, if operation \addModuleNoSpace[a,b,c] is executed in a step, than $c=a+b$ is expected in the following step. 
\Tool allows measuring the difference between the expected $c=a+b$, and actual simulation value.
To use the error evaluation, a user simply runs a Mathematica function we provide, called \textit{EvaluateError} which accepts two inputs: (1) \Tool program and (2) simulation duration. 
This helps users analyze the error, and detect if the error builds up over time.

We analyze the value of operand $a$ from GCD example
\figname~\ref{fig:gcd}, and plot the error in
\figname~\ref{fig:gcdError}. 
In \figname~\ref{fig:gcdError}, the x-axis represents time,
while the y-axis shows the difference between expected and actual value of $a$. 
Note that the error is sufficiently small that the algorithm executes correctly throughout the analyzed time.
The error is not constant, which opens
interesting questions of correlating the error with instructions in
the program. To correlate error with program instructions we examine
the GCD simulation (\figname~\ref{fig:gcdSimulation}). By looking at the time axis, it is easy to
connect the first two spikes of the error with the subtraction of $a$.

We provide the error evaluation framework with the vision of it being a guiding
element for programming in \Tool. We found this technique particularly
useful for validation of programs, analyzing the error, understanding
the sources of error, and redesigning the CRN for correctness.

\vspace{-5pt}
\section{Evaluation}
\label{sec:evaluation}
In this section, we evaluate \Tool on a set of discrete and real-valued examples. Later in this section, we characterize the error of basic modules.

\vspace{-10pt}
\subsection{Examples}
We first show a set of discrete algorithms implemented in \Tool---counter, factorial, integer division, integer square root;
followed by real-valued algorithms---approximating Euler's and $\pi$ constants.

\Subsubsection{Discrete Counter}

\begin{figure}[!t]
  \centering
  \begin{subfigure}[b]{0.45\textwidth}
\begin{lstlisting}[
  language=Mathematica,
  %
  %
  %
  escapechar=\#
]
crn = {
  conc[c,c0], conc[cInitial,c0],
  conc[one,1], conc[zero,0],
  step[{
    sub[c,one,cnext],
    cmp[c,zero]
  }],
  step[{
    ifGT[{ ld[cnext,c] }],
    ifLE[{ ld[cInitial,c] }]
  }]
}
\end{lstlisting}
     \caption{\Tool code.}
    \label{fig:counterCode}
  \end{subfigure}
  \hfill
  \begin{subfigure}[b]{0.45\textwidth}
    \includegraphics[width=\textwidth]{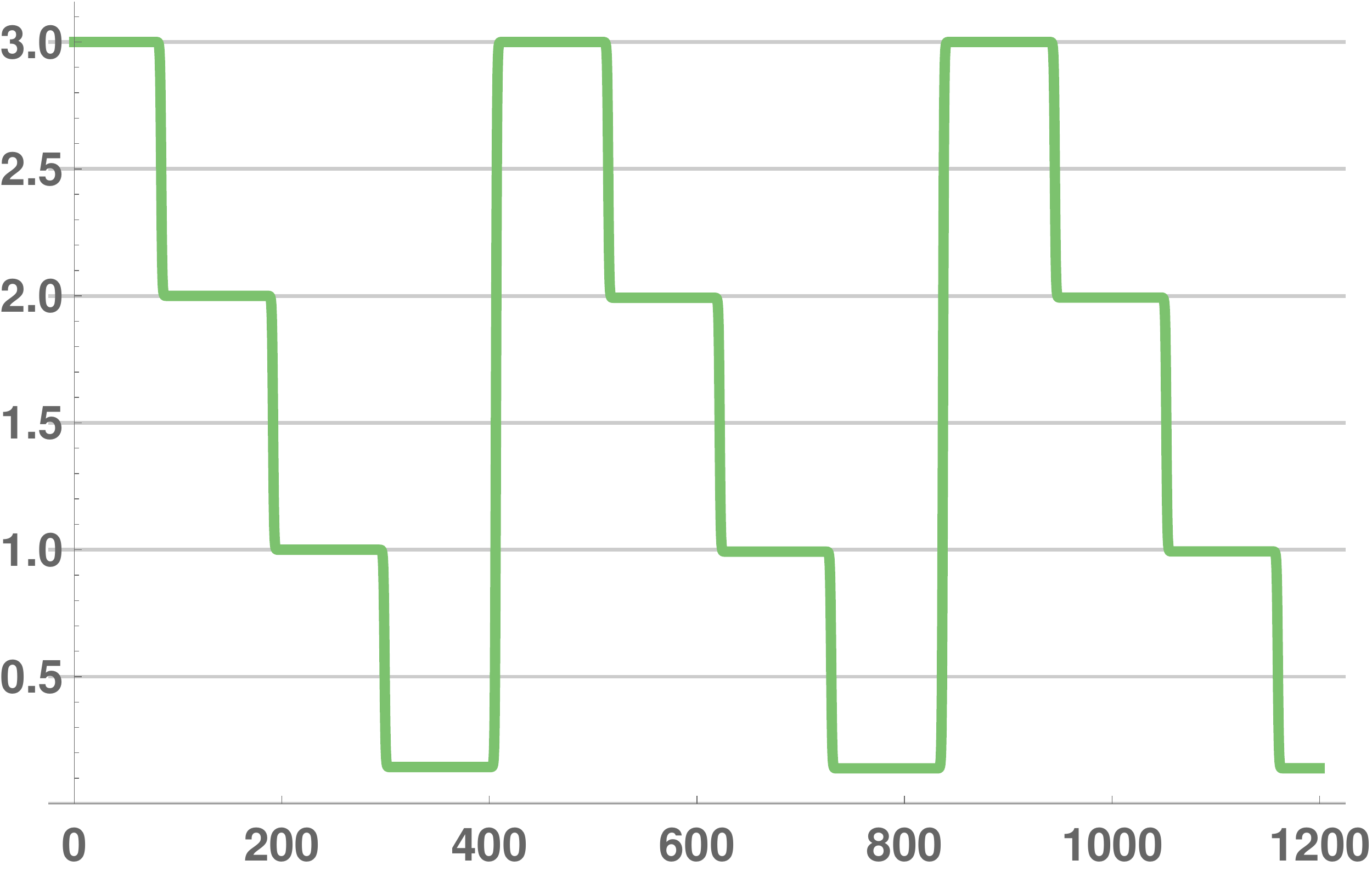}
    \caption{Simulation results for \mbox{$c0=3$}; value of $c$ is shown (green line).}
    \label{fig:counterSimulation}
  \end{subfigure}
  \caption{Discrete Counter.}
  \label{fig:counter}
  \vspace{-10pt}
\end{figure}

We implement a discrete counter that counts from a predefined value to zero, and repeats the process.
Fig~\ref{fig:counter} shows both the \Tool program and the simulation results.
Variable $c$ stores value of the counter, 
\emph{cInitial} stores the initial value of the counter for later refills,
while \emph{one} and \emph{zero} store constants $0$ and $1$, respectively.
Initial concentrations of the species are set in lines 2-3;
where \emph{c0} is a parameter of the program representing the initial value of the counter.
In step 1 (lines 4-7), \emph{one} is subtracted from the counter and stored into \emph{cnext}, and at the same time the counter is compared with \emph{zero}.
In step 2 (lines 8-11), if the counter is zero, then its value is reset to the initial value;
otherwise, \emph{cnext} is stored into the counter.
Steps exhibit looping behavior, thus described commands repeat.

\Subsubsection{Factorial}

\begin{figure}[!t]
  \centering
  \begin{subfigure}[b]{0.45\textwidth}
\begin{lstlisting}[
  language=Mathematica,
  %
  %
  %
  escapechar=\#
]
crn={
  conc[f,1], conc[one,1], conc[i,f0],
  step[{
    cmp[i,one],
    mul[f,i,fnext],
    sub[i,one,inext]
  }],
  step[{
    ifGT[{    
      ld[inext,i],
      ld[fnext,f]
    }]
  }]
}
\end{lstlisting}
     \caption{\Tool code.}
    \label{fig:factorialCode}
  \end{subfigure}
  \hfill
  \begin{subfigure}[b]{0.45\textwidth}
    \includegraphics[width=\textwidth]{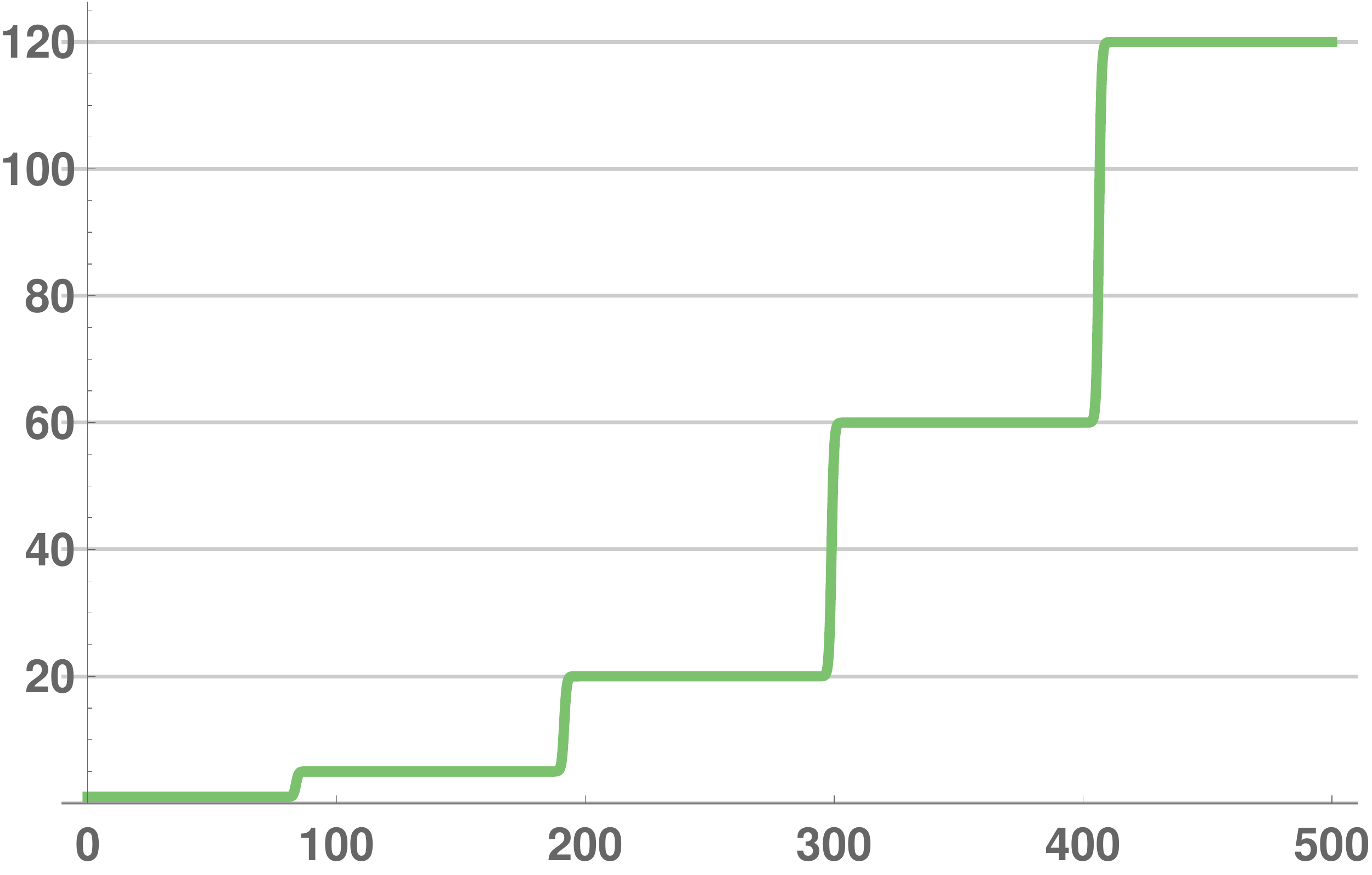}
    \caption{Simulation results for \mbox{$f0=5$}; value of $f$ is shown (green line).}
    \label{fig:factorialSim}
  \end{subfigure}
  \caption{Factorial.}
  \label{fig:factorial}
\end{figure}

We implement a program in \Tool that computes the factorial function.
Fig~\ref{fig:factorial} shows both the program and simulation results.
To compute the factorial of a positive integer $n$, we store $n$ in the iterator variable $i$,
and repeatedly multiply $f$ with $i$, decreasing $i$ until it becomes zero.
Initial concentrations of the species are defined in line 2.
In step 1 (lines 3-7), the iterator $i$ is compared with $one$ to check the termination condition,
$f$ is multiplied with $i$ storing the value in the temporary variable $fnext$,
and finally $i$ is decremented storing the value in the temporary $inext$.
In step 2 (lines 8-13), if $i > 1$, the temporary variables are stored back to $f$, and $i$.

\Subsubsection{Integer Division}

\begin{figure}[!t]
  \centering
  \begin{subfigure}[b]{0.45\textwidth}
\begin{lstlisting}[
  language=Mathematica,
  %
  %
  %
  escapechar=\#
]
crn={
  conc[a,a0], conc[b,b0], conc[one,1],
  step[{
    cmp[a,b]
  }],
  step[{
    IfGE[{
      sub[a,b,anext],
      add[q,one,qnext]
    }]
  }],
  step[{
    IfGE[{
      ld[anext,a],
      ld[qnext,q]
    }],
    ifLT[{ld[a,r]}]
  }]
};
\end{lstlisting}
     \caption{\Tool code.}
    \label{fig:divCode}
  \end{subfigure}
  \hfill
  \begin{subfigure}[b]{0.45\textwidth}
    \includegraphics[width=\textwidth]{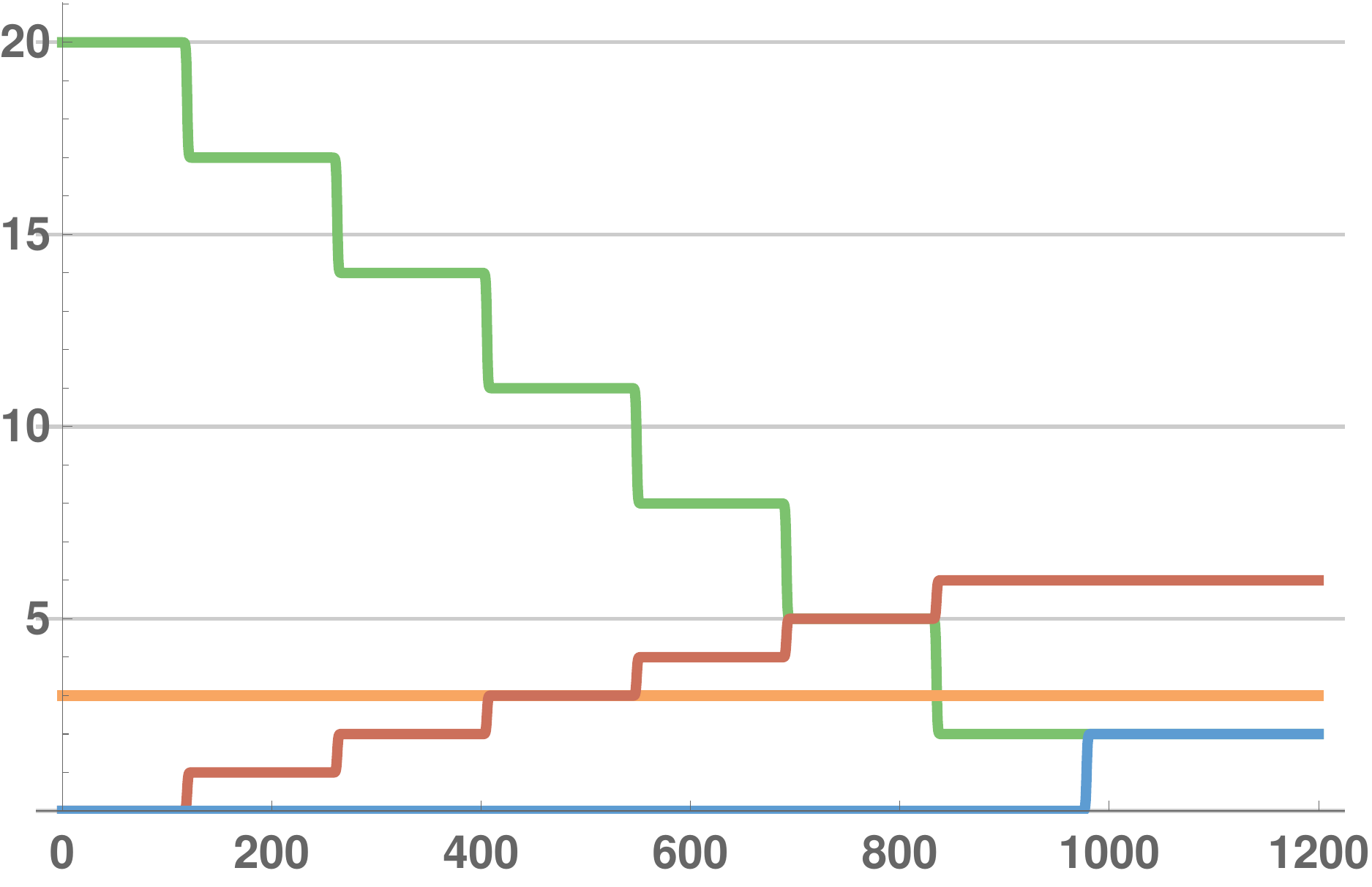}
    \caption{Simulation results for \mbox{$a0=20$}, \mbox{$b0=3$};
      values of $a$ (green), $b$ (orange), $q$ (red), and of $r$ (blue) are shown.}
    \label{fig:divSimulation}
  \end{subfigure}
  \caption{Division.}
  \label{fig:div}
  \vspace{-10pt}
\end{figure}

We implement integer division of two numbers, 
computing the quotient and the remainder of the operation.
Fig~\ref{fig:div} shows both the program and simulation results.
Variable $a$ stores the dividend, $b$ the divisor, $q$ the quotient, and $r$ the remainder.
The divisor is subtracted from the dividend until the dividend becomes smaller than the divisor.
In step 1 (lines 3-5), the dividend and divisor are compared to detect if the termination condition is satisfied.
In step 2 (lines 6-11), if $a>b$, the divisor is subtracted from the dividend, and the quotient is incremented.
In step 3, if $a>b$, the new values for the dividend and quotient are restored from the temporary variables into the original ones.
Also, in step 3, if $a<b$, the dividend is stored into the remainder (line 17).

\Subsubsection{Integer Square Root}

\begin{algorithm}
  \begin{algorithmic}[1]
    \Procedure{Int Sqrt}{$n$}
    \State $z\gets 0$
    \While{$(z+1)^2 \le n$}
    \State $z\gets z + 1$
    \EndWhile
    \State \textbf{return} $z$
    \EndProcedure
  \end{algorithmic}
  \caption{Integer square root.}
  \label{fig:intSqrtAlgorithm}
\end{algorithm}

\begin{figure}[!t]
  \centering
  \begin{subfigure}[b]{0.45\textwidth}
\begin{lstlisting}[
  language=Mathematica,
  escapechar=\#
]
crn = {
  conc[one,1], conc[n,n0],
  step[{ #\label{intSqrt:lineStep1Begin}#
    add[z,one,znext],
    mul[znext,znext,zpow],
    cmp[zpow,n]
  }], #\label{intSqrt:lineStep1End}#
  step[{ #\label{intSqrt:lineStep2Begin}#
    ifLT[{ld[znext,z]}],
    ifGE[{ld[z,out]}]
  }] #\label{intSqrt:lineStep2End}#
};
\end{lstlisting}
     \caption{\Tool code.}
    \label{fig:intSqrtCode}
  \end{subfigure}
  \hfill
  \begin{subfigure}[b]{0.45\textwidth}
    \includegraphics[width=\textwidth]{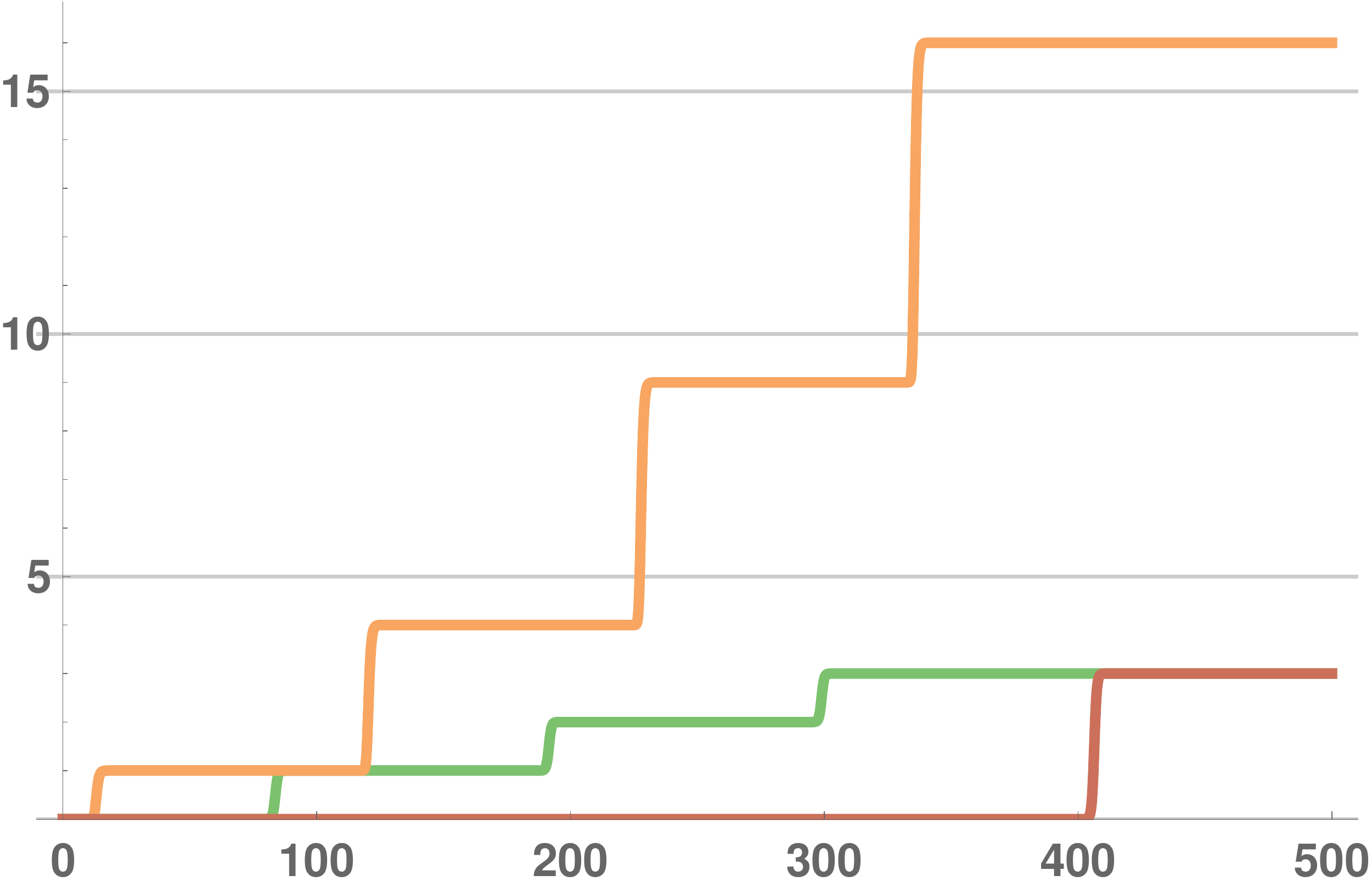}
    \caption{Simulation results for \mbox{$n0=10$}.
      Values of $z$ (green), $zpow$ (orange), and $out$ (red) are shown.}
    \label{fig:intSqrtSimulation}
  \end{subfigure}
  \caption{Integer square root.}
  \label{fig:intSqrt}
\end{figure}

We implement a program that finds the integer square root of a number.
\algname~\ref{fig:intSqrtAlgorithm} shows the pseudo-code algorithm:
the square root of a number $n$ is found by iterating through positive integer numbers: $0,1,2,$ etc, until square of the number overshoots $n$.
We map this algorithm to a \Tool program, and show the code and simulation results in \figurename~\ref{fig:intSqrt}.
In step 1 (lines \ref{intSqrt:lineStep1Begin}-\ref{intSqrt:lineStep1End}),
$z$ is incremented ($znext:=z+1$), and the square of $z+1$ is computed ($zpow:=znext*znext$) and compared with $n$.
In step 2 (lines \ref{intSqrt:lineStep2Begin}-\ref{intSqrt:lineStep2End}),
if $zpow<n$, $znext$ is stored into $z$; otherwise, the result is computed and stored in $out$.

\Subsubsection{Euler's number approximation}
\label{example:euler}

\begin{figure}[!t]
  \centering
  \begin{subfigure}[b]{0.45\textwidth}
\begin{lstlisting}[
  language=Mathematica,
  escapechar=\#
]
crn = {
  conc[e, 1], conc[element, 1], #\label{euler:lineConcBegin}#
  conc[divisor, 1], conc[one, 1],
  conc[divisorMultiplier, 1], #\label{euler:lineConcEnd}#
  step[{ #\label{euler:lineStep1Begin}#
    div[element, divisor, elementNext],
    add[divisor, one, divisorNext],
    add[e, elementNext, eNext]
  }], #\label{euler:lineStep1End}#
  step[{ #\label{euler:lineStep2Begin}#
    ld[elementNext, element],
    ld[divisorNext, divisor],
    ld[eNext, e]
  }] #\label{euler:lineStep2End}#
};
\end{lstlisting}
     \caption{\Tool code.}
    \label{fig:eulerCode}
  \end{subfigure}
  \hfill
  \begin{subfigure}[b]{0.45\textwidth}
    \includegraphics[width=\textwidth]{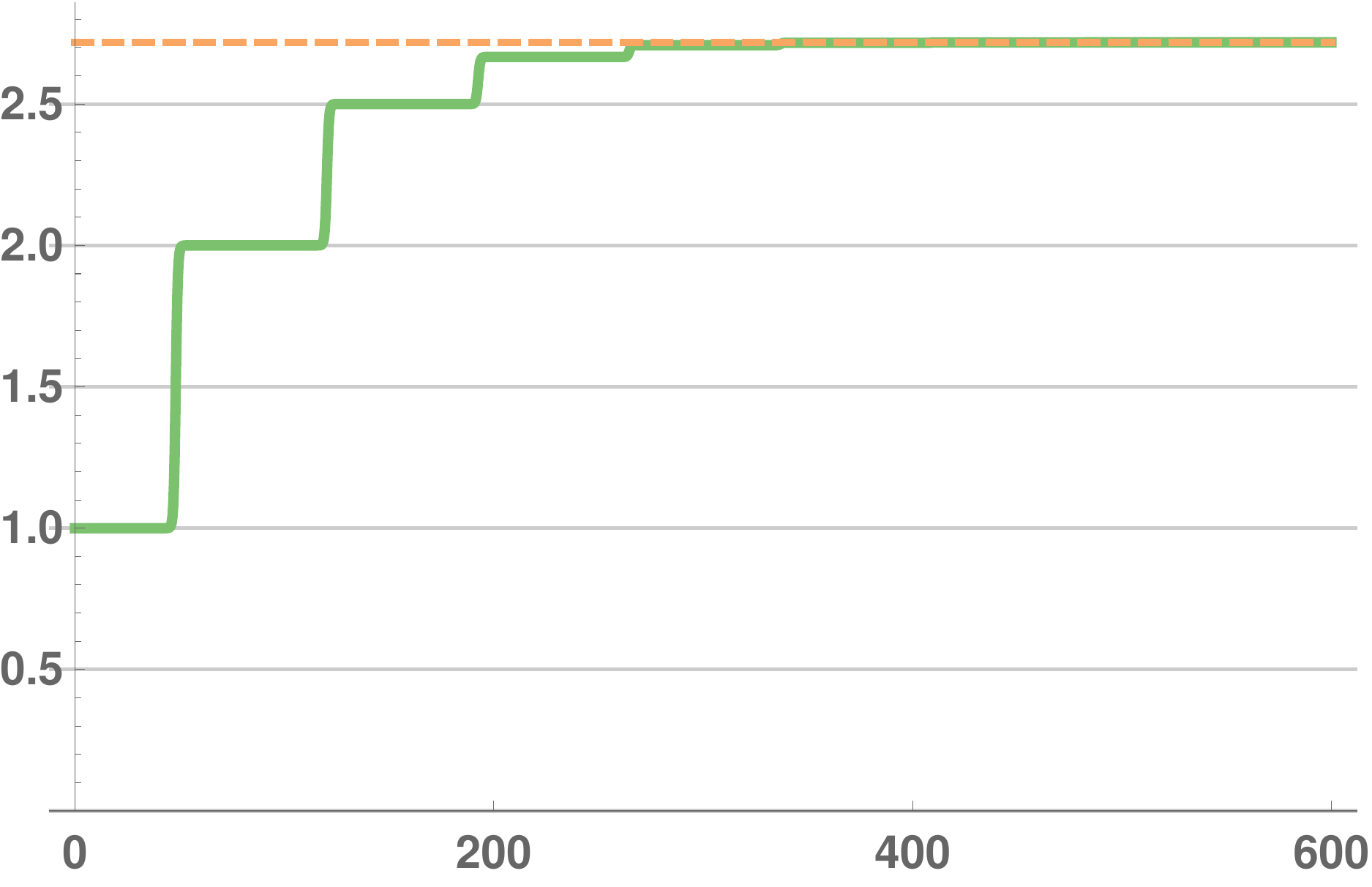}
    \caption{Simulation results. Approximation of Euler's number is
    shown in green line, while dashed orange line shows the
    correct value as a reference.}
    \label{fig:eulerSimulation}
  \end{subfigure}
  \caption{Approximating Euler's number through infinite series.}
  \label{fig:euler}
\end{figure}

So far we presented discrete algorithms, however chemistry naturally allows for real-valued (analog) computations.
For programming with real values we make use of \Tool module for performing division.
The \divModule module follows the same design principles and characteristics as other arithmetic modules we presented (see Table~\ref{table:modules}).

We implement a program that approximates Euler's number.
Euler's number can be computed using the following infinite series:
\[ e=\sum_{n=0}^{\infty} \frac{1}{n!} = \frac{1}{1} + \frac{1}{1} + \frac{1}{1 \cdot 2} + \frac{1}{1 \cdot 2 \cdot 3} + ... \]
We map this program into \Tool code, as shown in Fig~\ref{fig:euler}.
Variable $e$ stores the current approximation of the constant, while $element$ stores the current element of the series.
In step 1 (lines \ref{euler:lineStep1Begin}-\ref{euler:lineStep1End}),  $element$ is divided by the $divisor$,
$divisor$ is incremented for the next iteration, and $e$ is incremented by the current element of the series.
In step 2 (lines \ref{euler:lineStep2Begin}-\ref{euler:lineStep2End}),
the temporary variables $elementNext$, $eNext$, and $divisorNext$, are restored into the original variables.
The precision achieved at the end of simulation is up to 5 decimal digits---2.71828.

\Subsubsection{Approximating $\pi$}
\label{example:pi}

\begin{figure}[!t]
  \centering
  \begin{subfigure}[b]{0.45\textwidth}
\begin{lstlisting}[
  language=Mathematica,
  escapechar=\#
]
crn={
  conc[four, 4],
  conc[divisor1, 1],
  conc[divisor2, 3],
  conc[pi, 0],
  #\stepKeywordNoSpace#[{ #\label{pi:lineStep1Begin}#
    #\divModuleNoSpace#[four, divisor1, factor1],
    #\addModuleNoSpace#[divisor1, four, divisor1Next],
    #\divModuleNoSpace#[four, divisor2, factor2],
    #\addModuleNoSpace#[divisor2, four, divisor2Next],
    #\subModuleNoSpace#[factor1, factor2, factor],
    #\addModuleNoSpace#[pi, factor, piNext]
  }], #\label{pi:lineStep1End}#
  #\stepKeywordNoSpace#[{ #\label{pi:lineStep2Begin}#
    #\ldModuleNoSpace#[divisor1Next, divisor1],
    #\ldModuleNoSpace#[divisor2Next, divisor2],
    #\ldModuleNoSpace#[piNext, pi]
  }] #\label{pi:lineStep2End}#
};
\end{lstlisting}
     \caption{\Tool code.}
    \label{fig:piCode}
  \end{subfigure}
  \hfill
  \begin{subfigure}[b]{0.45\textwidth}
    \includegraphics[width=\textwidth]{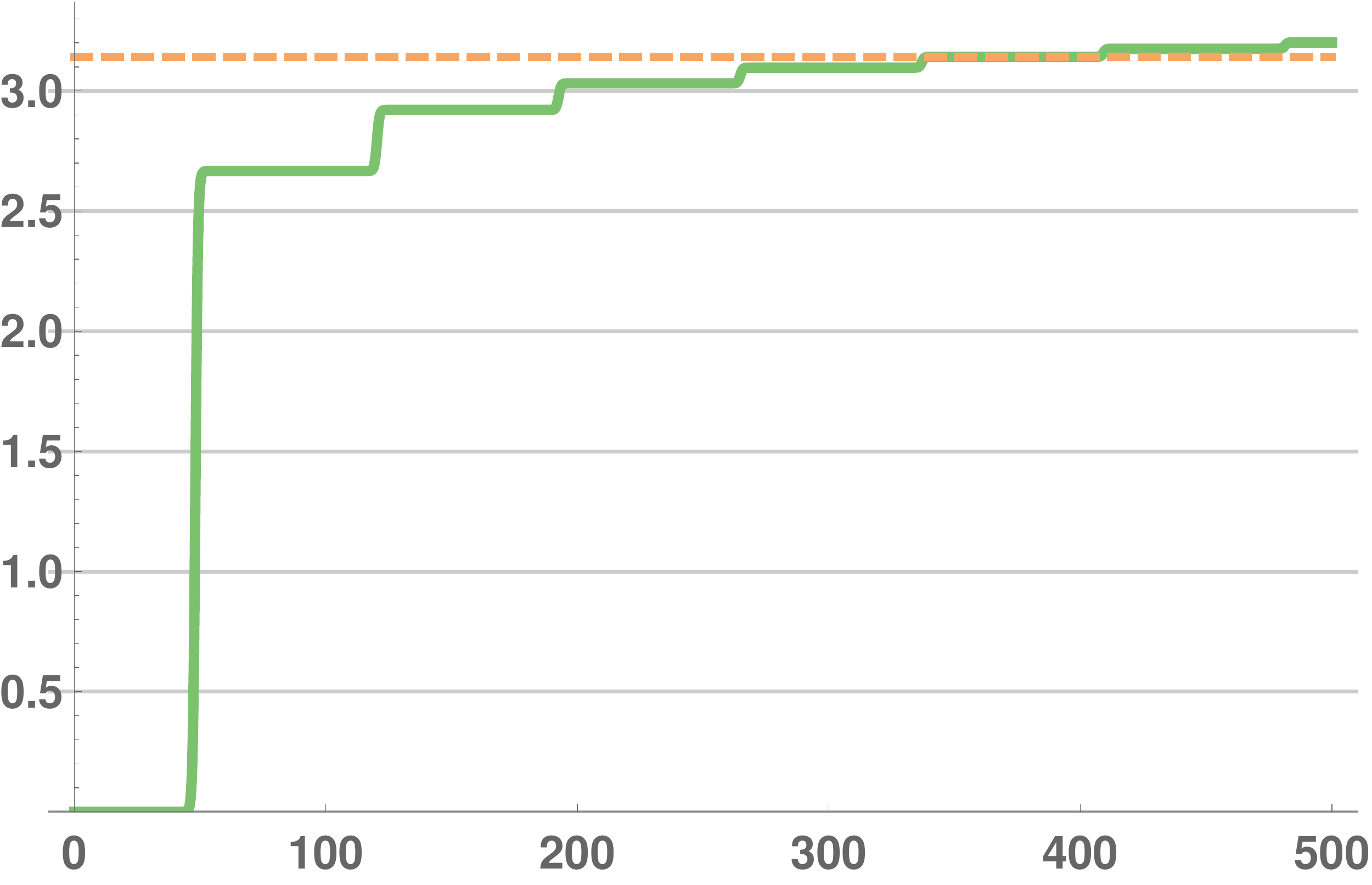}
    \caption{Simulation results. Approximation of $\pi$ constant is
    shown in green line, while dashed orange line shows the
    correct value as a reference.}
    \label{fig:piSimulation}
  \end{subfigure}
  \caption{Approximating Pi constant through infinite series.}
  \label{fig:pi}
\end{figure}

We approximate the constant $\pi$ via a \Tool program.
We rely on the following infinite series to do so:
\[ \pi = \frac{4}{1} - \frac{4}{3} + \frac{4}{5} - \frac{4}{7} + \frac{4}{9} - \frac{4}{11} + ... \]
Fig~\ref{fig:pi} shows both the code and simulation.
In step 1 (lines \ref{pi:lineStep1Begin}-\ref{pi:lineStep1End}) following instructions are executed:
(a) $4$ is divided by the current divisor \emph{divisor1} and stored into \emph{factor1},
(b) $4$ is divided by the \emph{divisor2} and stored into \emph{factor2},
(c) \emph{factor1} subtracted by \emph{factor2} is added to the $pi$,
(d) \emph{divisor1} and \emph{divisor2} increased by $2$ are stored into \emph{divisor1Next} and \emph{divisor2Next}, respectively.
In step 2 (lines \ref{pi:lineStep2Begin}-\ref{pi:lineStep2End}),
the temporary variables \emph{divisor1Next}, \emph{divisor2Next}, and \emph{piNext} are restored to the original variables.
At the end of simulation we measure the output value $3.20185$. %
Error evaluation shows that the error is greater in computing $\pi$ compared to Euler's number due to using subtraction (of close values),
which is the most error-prone out of all arithmetic operations we present (see~\ref{sec:errorCharacterization}).

\Subsubsection{Size of CRNs}

\begin{table}[!t]
\begin{center}
\begin{tabular}{l | c | c}
\toprule
\HighlightCell{Program} & \HighlightCell{\#Species} & \HighlightCell{\#Reactions} \\
\midrule
Discrete Counter & $25$ & $31$ \\
Factorial & $26$ & $33$ \\
Integer Division & $32$ & $39$ \\
Integer Square Root & $26$ & $32$ \\
Euler & $24$ & $20$ \\
$\pi$ & $29$ & $29$ \\
\bottomrule
\end{tabular}
\end{center}
\caption{Size of CRNs.}
\label{table:crn-size}
\end{table}

Table~\ref{table:crn-size} shows the sizes of CRNs, in terms of the number of reactions and species, that result from compilation of corresponding \Tool programs.

\subsection{Error Characterization}
\label{sec:errorCharacterization}
In this section we evaluate error of the \Tool modules.

\Subsubsection{Error of Arithmetic Modules}
\label{appendix:error:modules}

\begin{figure}[!t]
  \centering
  \begin{subfigure}[b]{0.42\textwidth}
    \includegraphics[width=\textwidth]{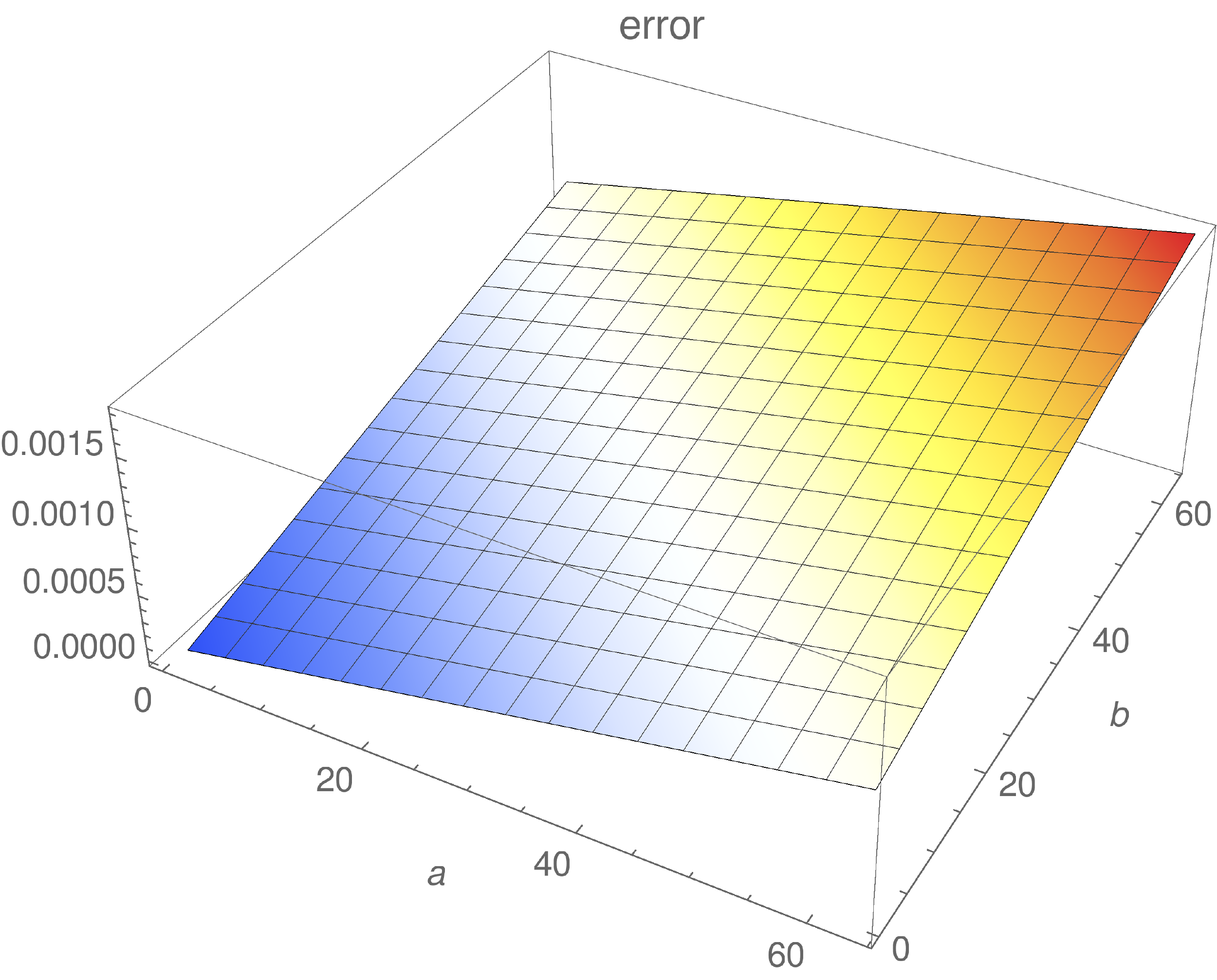}
    \caption{Add evaluation.}
    \label{fig:addEval}
  \end{subfigure}
  \hfill
  \begin{subfigure}[b]{0.42\textwidth}
    \includegraphics[width=\textwidth]{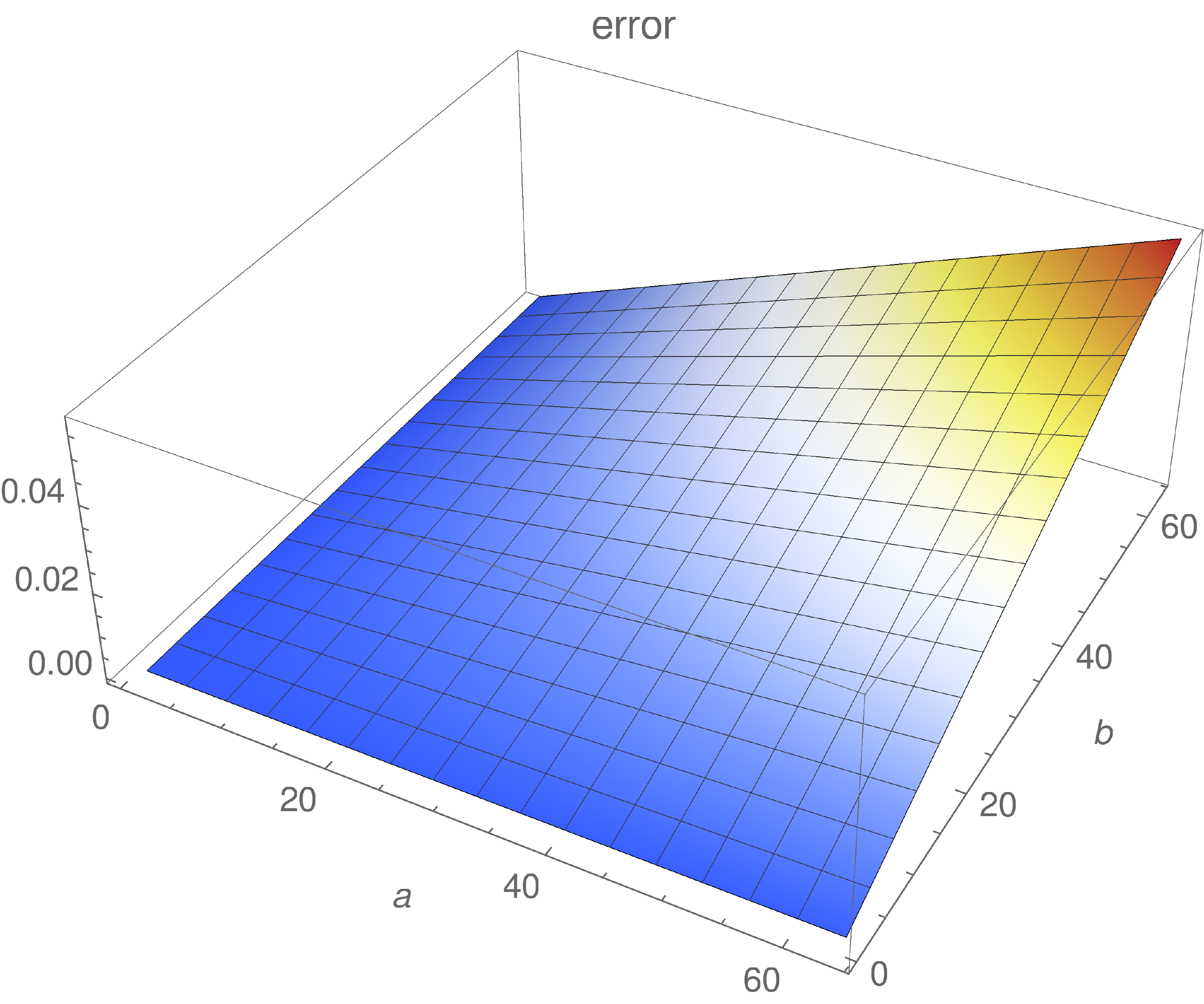}
    \caption{Mul evaluation.}
    \label{fig:mulEval}
  \end{subfigure}
  \hfill
  \begin{subfigure}[b]{0.42\textwidth}
    \includegraphics[width=\textwidth]{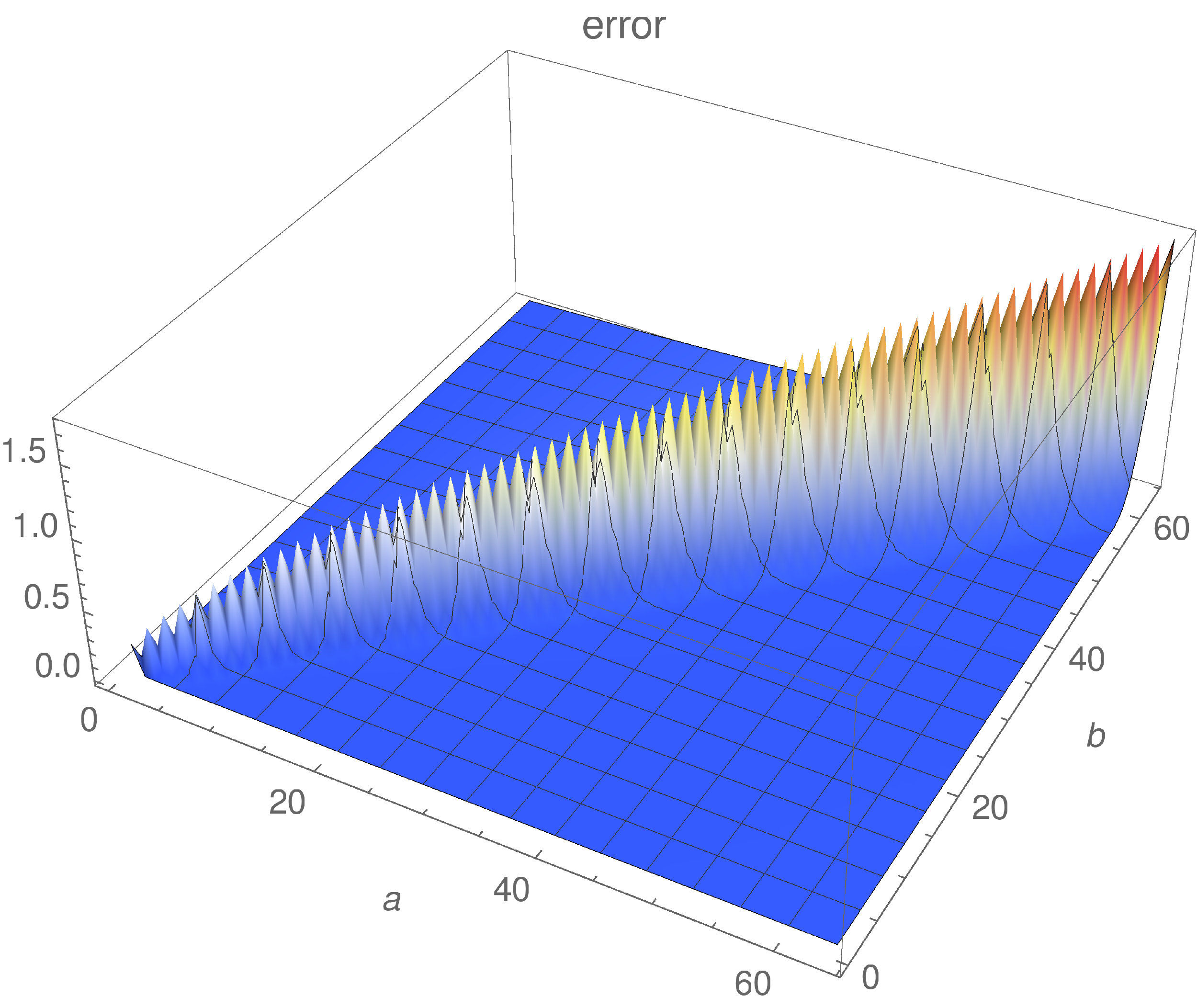}
    \caption{Sub evaluation.}
    \label{fig:subEval}
  \end{subfigure}
  \hfill
  \begin{subfigure}[b]{0.42\textwidth}
    \includegraphics[width=\textwidth]{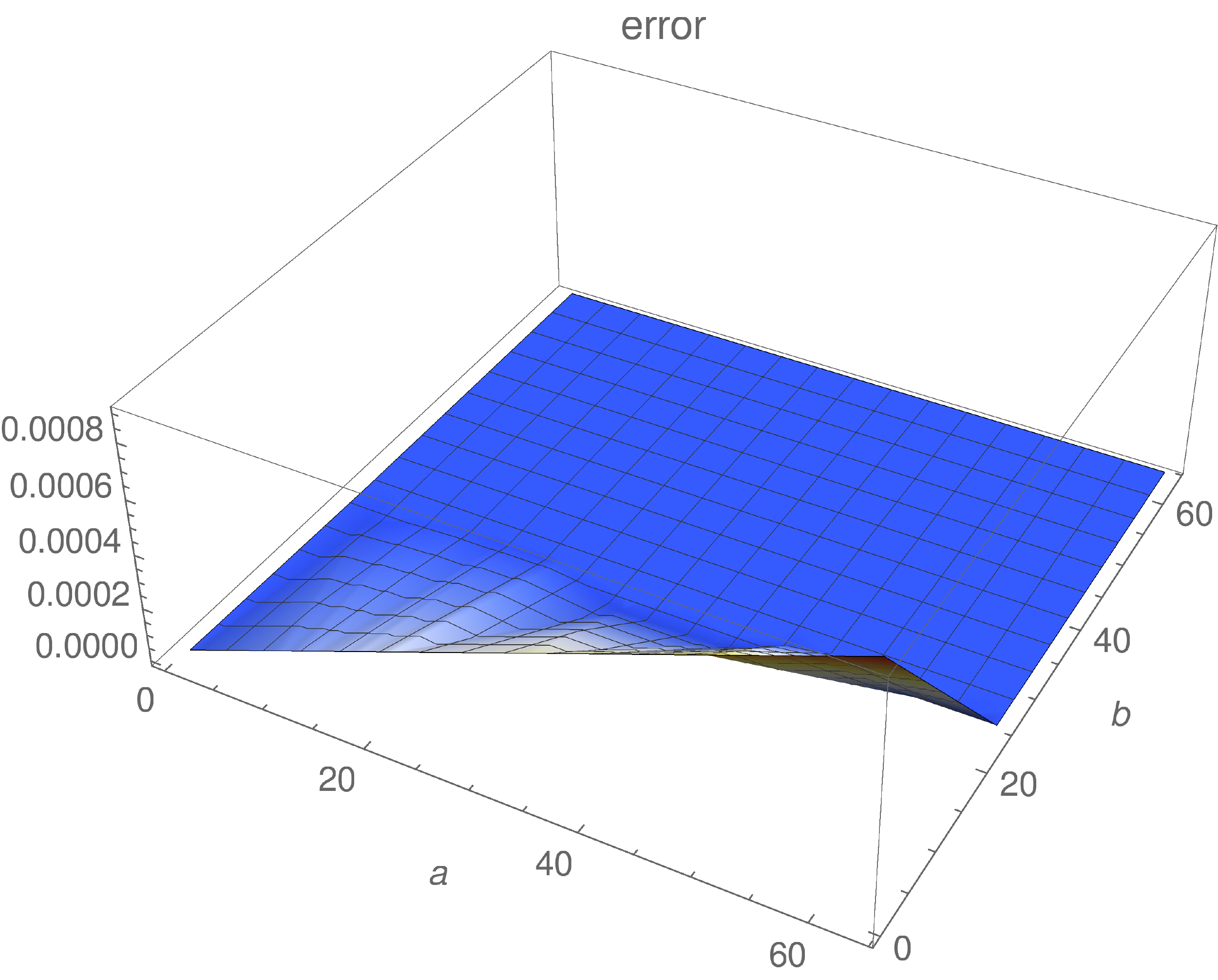}
    \caption{Div evaluation.}
    \label{fig:dvdEval}
  \end{subfigure}
  \caption{Error evaluation of arithmetic modules.
    Axes $a$ and $b$ show the values of the first and second operand, respectively;
    the height shows the absolute value of the error
    (difference between the correct and actual value of the operation).}
  \label{fig:modulesErrorEvaluation}
\end{figure}

Using our error evaluation mechanisms (section~\ref{sec:technique:errorEval}) we analyze the error of the modules.
We evaluate each module separately, on different inputs, to characterize its behavior.
\figname~\ref{fig:modulesErrorEvaluation} shows the error evaluation results,
where $a$ and $b$ axes reflect the values of the first and second operand, respectively, and the height shows the magnitude of the error.
The plots provide useful information such as:
(a) The error of the \mulModule and \addModule modules increases as the value being computed increases;
(b) Since these CRNs are symmetric with respect to the inputs, the error does not depend on the order of the arguments;
(c) The \subModule module exhibits the largest error when the inputs are close to each other,
and in general, has a higher error than the other arithmetic modules.
This information is useful when designing \Tool programs:
error-prone subtraction of inputs close to each other is the reason why the error in the program approximating $\pi$ (\ref{example:pi}) is higher than in the one approximating Euler's number (\ref{example:euler}).
Having this in mind, a user can optimize a program;
for example, the subtraction of close operands can often be done in alternative, less error-prone ways (see below, \figname~\ref{fig:subAlternativeCode}).
We plan to add runtime assertions to \Tool programs that alert for possible issues in the program;
for example, when values being subtracted are closer than $\epsilon$ to each other.

\Subsubsection{Reducing the error through program refactoring}
\label{appendix:error:refactoring}

\begin{figure}[!t]
  \begin{subfigure}[b]{0.45\textwidth}
    \includegraphics[width=\textwidth]{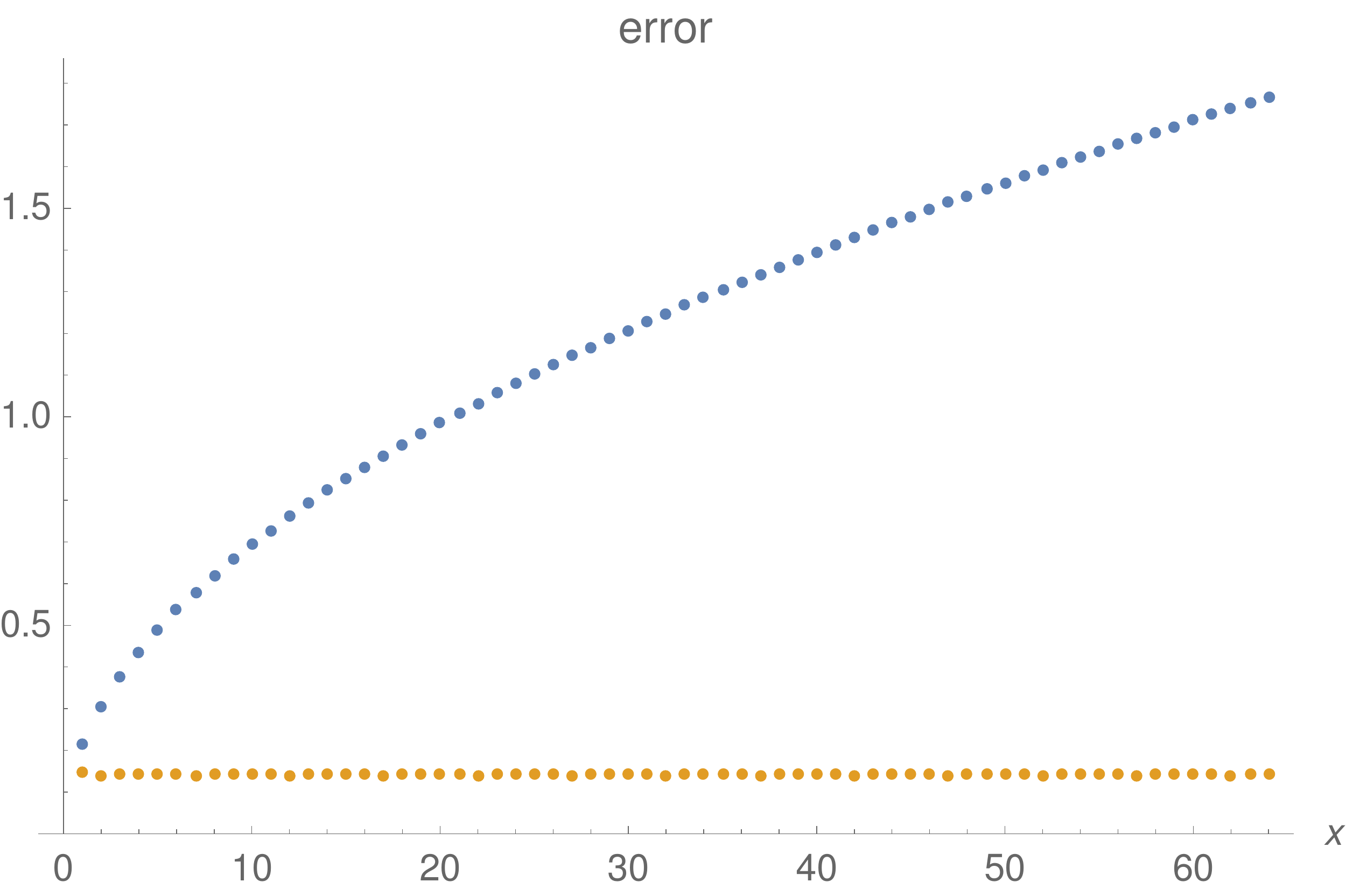}
    \caption{Comparing error of Sub module (blue lines) and
      alternative way to subtract (orange lines). X-axis show the
      value of both minuend and subtrahend.}
    \label{fig:subAlternativeEval}
  \end{subfigure}
  \hfill
  \begin{subfigure}[b]{0.45\textwidth}
\begin{lstlisting}[
  language=Mathematica,
  %
  %
  %
  escapechar=\#
]
crn = {
  conc[a, a0], conc[b, b0],
  conc[one, 1], conc[zero, 0],
  step[{
    cmp[b, zero]
  }],
  step[{
    ifGE[{
      sub[a, one, anext],
      sub[b, one, bnext]
    }]
  }],
  step[{
    ifGE[{
      ld[anext, a],
      ld[bnext, b]
    }]
  }]
}
\end{lstlisting}
     \caption{Alternative way to subtract.}
    \label{fig:subAlternativeCode}
  \end{subfigure}
  \caption{Comparing error of \subModule with the alternative way of subtracting (\figname~b).
    Error evaluation is shown (\figname~a) for the cases when the operands are equal (minuend and subtrahend same),
    since \subModule exhibits the highest error in that case.}
  \label{fig:subAlternative}
\end{figure}

The \subModule module has a high error when the operands are close to each other. 
In this section we show an example where the error can be reduced by replacing the \subModule module with an alternative subtraction algorithm.
\figname~\ref{fig:subAlternativeCode} shows the alternative code for performing subtraction.
The value of $b$ is subtracted from $a$, by repeatedly subtracting $1$ from both $a$ and $b$, until $b$ reaches $0$.
This method ensures smaller error which is also constant in time, however it is less time efficient.

\newpage
\section{Related Work}
\label{sec:related:work}

\textbf{Computational power of chemical reaction networks.} 
Previous research demonstrated techniques of achieving complex behaviors in mass-action chemistry,
such as computing algebraic functions and
polynomials~\cite{BuismanETAL09ComputingAlgebraicFunctionsInCRNs,SalehiETAL17CRNsForComputingPolynomials,salehi2018computing},
logarithms~\cite{chou2017chemical},
implementing logic gates and finite state machines~\cite{hjelmfelt1992chemical,Magnasco97ChemicalKineticsIsTuringUniversal,ge2016formal}, and neural networks~\cite{hjelmfelt1991chemical,salehi2018computing}.
Moreover, the Turing completeness of chemistry has been proven,
using the strategy of implementing polynomial ODEs
(which have been previously shown to be Turing universal)
in mass-action chemical kinetics~\cite{FagesETAL17TuringCompletenessOfContinuousCRNs}.
Even though Turing complete, this translation to chemistry can result in infeasibly complex chemical reaction networks,
which motivates other, more direct methods. 

\textbf{Modular Reactions.} 
Adding even a single reaction to a CRN can completely change its dynamics,
which makes the design process challenging.
The idea of ``composable" reactions seeks a set of reactions that can be composed in a well-defined manner to implement more complex behaviors.
Buisman et al.~\cite{BuismanETAL09ComputingAlgebraicFunctionsInCRNs}
computed algebraic expressions by designing the core modules that implemented basic arithmetic operations,
which can be further composed to achieve more complex tasks.
Our goal is to make modular designs,
and we follow some of the proposed design principles for achieving the goal,
such as input-preserving CRNs.

\textbf{Synchronous computation}.
Previous work utilized synchronous logic to achieve complex tasks.
Soloveichik et al.\ implemented state machines in chemistry by relying on a ``rock-paper-scissors'' type of chemical clock (oscillator)~\cite{SoloveichikETAL10DNAUniversalSubstrate}.
We use the same clock module, 
with clock species acting catalytically to order reactions.
Jiang et al.~\cite{JiangETAL11SynchronousSequentialComputationsWithMolecularReactions}, also relying on clock species,
designed a model of memory in chemistry to support sequential computation,
demonstrating their technique on examples of a binary counter and a fast Fourier transform (FFT).
Previous work shows the promise of programming synchronous logic in reactions,
which we advance by providing an explicit programming language and framework for designing and testing wide-range of programs.

\textbf{Asynchronous computation.}
Huang et al.~\cite{HuangETAL12CompilingControlFlowIntoBiochemicalReactions} used ``absence indicators'' to implement complex algorithms such as integer division and GCD.
An absence indicator is a species that is present in high concentration when a target species is present in low concentration~\cite{SenumRiedel11ConstructsForChemicalComputation}.
Absence indicators can be used to drive a reaction when a particular reaction has finished, providing a method for executing modules in desired order.
Generally speaking, the absence indicator for species $A$ is produced at a constant rate and gets degraded by $A$.
The absence indicator has to be produced slowly, or else it will be present in non-negligible concentration even if $A$ is present.
The absence indicators in the literature relied on a difference between rate constants of several orders of magnitude; 
e.g.,~\cite{HuangETAL12CompilingControlFlowIntoBiochemicalReactions} uses two reaction rates, `fast' and `slow', where the fast rate needs to be orders (2-3) of magnitude larger to ensure the proper function of the system.
Since, in practice, biochemical systems allow for a restricted range of reaction rates, requiring a large difference in rates slows down the computation when the computation speed is dictated by the slow rates.
In contrast, we allow all reactions to take the same (or comparable) rate constants.

While the goal of our work is not to compare asynchronous and synchronous computation,
we mention a few insights 
which we gained through empirical studies.
First, absence indicators are not robust, and typically require fine tuning to get the system right.
Second, error detection is easier with synchronous logic.
Since all operations follow the clock signal, there is a direct mapping from a time moment to a command that is executing,
which provides a way to check correctness of a system at any point of time.
We provide a framework for implementing molecular programs
which is easily extensible, and can be used to compare synchronous and asynchronous logic.
We include support for absence indicators through the \ifAbsentKeyword construct,
thus allowing easier comparison of the two paradigms.

\section{Discussion and conclusions}
\label{sec:discAndConcl}
There are multiple ways in which we can further improve \Tool.
Note that currently every high-level module is mapped to exactly one CRN implementing the operation.
Letting the tool decide which implementation to use in different contexts could boost the performance. 
For example, the described modules have a useful property of preserving inputs,
but that property might not be needed in every case.
If the input preserving property is redundant,
\Tool could choose to use the more optimized version (for example the more compact subtraction CRN discussed above).
Also, we could provide a more flexible programming experience by
(a) letting the compiler automatically schedule instructions to different steps (instead of the explicit \texttt{step} construct);
(b) allowing the same species as both input and output of a module and automatically generate the additional instructions.

We plan to further explore the support for nested loops in \Tool.
Currently nested loops can be mimicked through conditional execution: the  loop condition is computed through comparison and the main loop conditionally executes the instructions of the desired loop.
Besides explicit support for nested loops,
future work will support nested conditionals by adding multiple flag species for multiple comparisons.

An important direction for future research concerns reducing the error in our construction, and understanding how it builds up over time.
We noticed that different algorithms, even computing the same function, accumulate varying levels of error.
For example, as seen in~\ref{appendix:error:modules}, the error of the Sub module increases with the magnitude of the operands, and also increases the closer they are.
However, we also found an alternative way to subtract, that keeps the error constant and independent of the operands (see \figname~\ref{fig:subAlternativeCode}) at the cost of a slower run-time.

Our error analysis shows that for most examples we tried, but not all, error builds up over the course of the computation. 
For \Tool programs where the error builds up in this way, there is some maximum input complexity beyond which the error overwhelms the output. 
Can all \Tool programs be refactored (preferably automatically) to bound the cumulative error of every module such that it does not build up over time?
Note that if this were possible, we would obtain another, more efficient, way to achieve Turing universality.

To the best of our knowledge we are the first to provide an imperative programming language which compiles to chemical reaction networks.
Moreover, we build tools that can help users get a better understanding of CRNs and improve their design.
Although absolutely correct computation is not achieved, we provide tools that help understand why error occurs, and thus help improve the design of CRNs.
We release our toolkit as open-source, to encourage new research and improvement of the \Tool, with the hope of advancing the engineering of information processing molecular systems.

In this work we advance the state of imperative programming with CRNs. 
It remains an important open question, however, whether fundamentally new programming paradigms are needed to fully and effectively utilize the computational power of chemistry.
 
\noindent
\textbf{Acknowledgments.}
We thank the fellow students of EE 381V (Programming With Molecules)
at The University of Texas at Austin for constructive discussions on the material presented in this paper.
We also thank Keenan Breik, Cameron Chalk, Milos Gligoric, Aleksandar Milicevic, Boya Wang and Kaiyuan Wang for their feedback on this work.
This research was partially supported by the US National Science Foundation
under Grants Nos.~CCF-1618895, CCF-1718903, CCF-1652824, and CCF-1704790.

\bibliographystyle{splncs03}
\bibliography{citing/conf,citing/bib}

\end{document}